\documentclass[12pt]{article}
\pdfoutput=1 % for JHEP to accept pdf images with graphix

\usepackage{amsmath}
\usepackage{amssymb}
\usepackage{amsfonts}
\usepackage{graphicx}           % use \pdfoutput=1 after \documentclass
\usepackage[utf8]{inputenc}     % inspire bibs
\usepackage{aas_macros}				  % ads bibs
\usepackage{bm}                 % \boldsymbol
\usepackage{amsthm}
  \newtheoremstyle{mystyle}%    % Name
  {1em}%                           % Space above
  {1em}%                           % Space below
  {\itshape}%                   % Body font
  {}%                           % Indent amount
  {\bfseries}%                  % Theorem head font
  {.}%                          % Punctuation after theorem head
  { }%                          % Space after theorem head, ' ', or \newline
  {}%                           % Theorem head spec (can be left empty, meaning `normal')

\theoremstyle{mystyle}

\usepackage{slashed}				% \slashed{k}
\usepackage{mathrsfs}			 	% Weinberg-esque letters
\usepackage{bbm}					  % \mathbbm{1} conflict: XeLaTeX 
\usepackage{cancel}		
\usepackage[normalem]{ulem} 		% for \sout
\usepackage{lineno}
\usepackage[dvipsnames]{xcolor}
\usepackage[hang,flushmargin]{footmisc} % no footnote indent

\usepackage{fancyhdr}		% preprint number
\usepackage{lipsum}			% block of text 
\usepackage{framed}			% boxed remarks
\usepackage{subcaption}	% subfigures
\usepackage{cite}			  % group cites
\usepackage{xspace}			% macro spacing

\usepackage{booktabs}		% professional tables
\usepackage{nicefrac}		% fractions in tables,
\usepackage[font=small]{caption} % caption font is small
\usepackage{float}         % for strict placement e.g. [H]

\usepackage[margin=2cm]{geometry}   % margins
\graphicspath{{figures/}}			% figure folder
\numberwithin{equation}{section}    % set equation numbering

\usepackage{multicol}
\usepackage{etoolbox}
\usepackage{relsize}
\patchcmd{\thebibliography}
  {\list}
  {\begin{multicols}{2}\smaller\list}
  {}
  {}
\appto{\endthebibliography}{\end{multicols}}

\let\oldenumerate\enumerate
\renewcommand{\enumerate}{
  \oldenumerate
  \setlength{\itemsep}{2pt}
  \setlength{\parskip}{0pt}
  \setlength{\parsep}{0pt}
}

\let\olditemize\itemize
\renewcommand{\itemize}{
  \olditemize
  \setlength{\itemsep}{1pt}
  \setlength{\parskip}{0pt}
  \setlength{\parsep}{0pt}
}

\renewcommand{\vec}[1]{\mathbf{#1}} % vectors are boldface
\definecolor{nirmal}{rgb}{1,0,1}
\definecolor{nirmalcomment}{rgb}{.4,.5,1}

\newcommand{\email}[1]{\href{mailto:#1}{#1}}
\newenvironment{institutions}[1][2em]{\begin{list}{}{\setlength\leftmargin{#1}\setlength\rightmargin{#1}}\item[]}{\end{list}}

\usepackage[
	colorlinks=true,
	citecolor=green!50!black,
	linkcolor=NavyBlue!75!black,
	urlcolor=green!50!black,
	hypertexnames=false]{hyperref}

\fancypagestyle{firststyle}{
	\rhead{\footnotesize%
	\texttt{LAPTH-001/23}%
	}}

\usepackage{etoc}

\newcommand{\mycomment}[1]{}
\usepackage{soul}

\newcommand{\orcid}[1]{\href{https://orcid.org/#1}{\includegraphics[width=8pt]{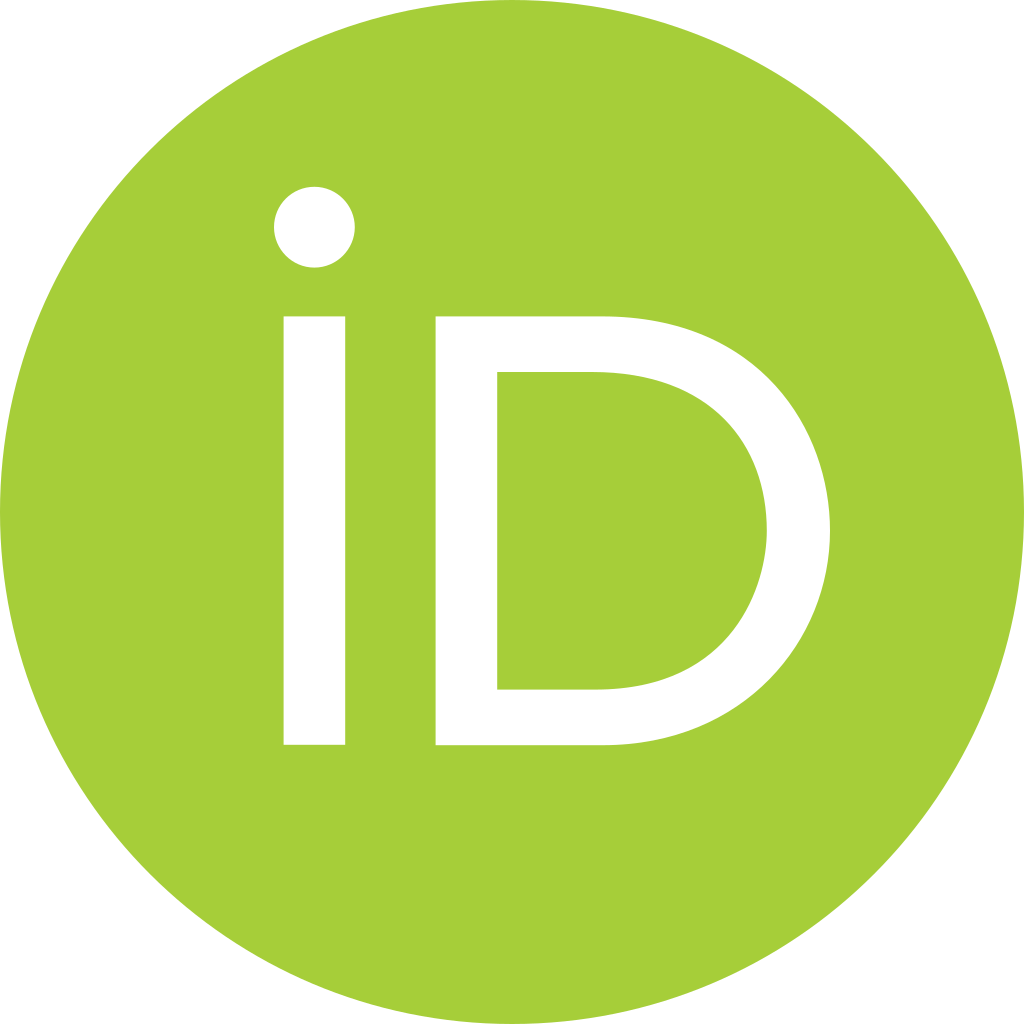}}}

\begin{document}
  \thispagestyle{firststyle}          %% Includes preprint #

\begin{center} 

    {\Large \bf Heating Neutron Stars with Inelastic Dark Matter \\ and Relativistic Targets }
    %\\
    %{\large Electrons Capture Dark Matter in Neutron Stars}

    \vskip .7cm

   { \bf 
    Gerardo Alvarez \orcid{0000-0003-4818-1501}$^{a}$,
    Aniket Joglekar \orcid{0000-0002-3572-6683}$^{b,c}$,
    Mehrdad Phoroutan-Mehr \orcid{0000-0002-9561-0965}$^{a}$,\\
    and
    Hai-Bo Yu \orcid{0000-0002-8421-8597}$^{a}$
    } 
   \\ 
   \vspace{.2em}
   { \tt \footnotesize
      \email{galva012@ucr.edu},
      \email{aniket@iittp.ac.in},
      \email{mphor001@ucr.edu},
      \email{haiboyu@ucr.edu}
    }
  
   \vspace{-.2cm}

   \begin{institutions}[1.3cm]
   \footnotesize
   $^{a}$ 
  {\it 
       Department of Physics \& Astronomy, University of California, Riverside, CA 92521, USA
  }
  \\ 
  \vspace*{0.05cm}   
  $^{b}$ 
   {\it 
      Department of Physics, Indian Institute of Technology Tirupati, Yerpedu Post, AP, 517619, India 
   }    
  \\ 
  \vspace*{0.05cm}   
  $^{c}$ 
  {\it 
       LAPTh, CNRS, Universit\'e Savoie Mont Blanc, F-74940 Annecy, France
  }
  \end{institutions}

\end{center}

%\date{\today}

\begin{abstract}
The dense environment of neutron stars makes them an excellent target for probing dark matter interactions with the Standard Model. We study neutron star heating from capture of inelastic dark matter, which can evade direct detection constraints. We investigate kinematics of the inelastic scattering process between quasirelativistic dark matter particles and ultrarelativistic targets in neutron stars, and derive analytical expressions for the maximal mass gap allowed for the scattering to occur. We implement them into a fully relativistic formalism for calculating the capture rate and apply it to various scenarios of inelastic dark matter. The projected constraints from neutron stars can systematically surpass those from terrestrial searches, including direct detection and collider experiments. Neutron stars can also be sensitive to the parameter space of inelastic self-interacting dark matter. Our results indicate that extreme astrophysical environments, such as neutron stars, are an important target for searching dark matter. 

\end{abstract}

\small
\setcounter{tocdepth}{2}
\tableofcontents
\normalsize
\clearpage

%\maketitle

\section{Introduction}
\label{sec:1}

Dark matter makes up more than $80\%$ of the mass in the universe, yet its nature is largely unknown. There has been growing interest in the search for dark matter signals via compact astrophysical objects~\cite{Goldman:1989nd,Gould:1989gw,
	Kouvaris:2007ay,
	Bertone:2007ae,
	McCullough:2010ai, 
	Kouvaris:2010jy,
	deLavallaz:2010wp, 
	Kouvaris:2010vv, 
	Kouvaris:2011fi, 
	McDermott:2011jp, 
	Guver:2012ba, 
	Bertoni:2013bsa, 
	Bramante:2013hn, 
	Bramante:2013nma, 
	Bell:2013xk, 
	Perez-Garcia:2014dra,
	Graham:2015apa,
	Bramante:2015cua,
	Cermeno:2016olb,
	Krall:2017xij, 
        McKeen:2018xwc,
	Graham:2018efk, 
        Gresham:2018rqo,
        Janish:2019nkk, 
        % Bell:2019pyc, %% leptons
	Camargo:2019wou,
	Dasgupta:2019juq,
        Garani:2019rcb,
        Ilie:2020nzp,
        Leane:2020wob,
        Acevedo:2020gro,
        Leane:2021ihh,
        Leane:2021tjj,
        Maity:2021fxw,
        Bell:2021fye,
        McKeen:2021jbh,
        Ilie:2021iyh,
        Ilie:2021umw,
        Bose:2021yhz,
        Bramante:2021dyx,
        DeRocco:2022rze,
        Collier:2022cpr,
        %Leane:2022hkk,
        Bramante:2022pmn,
        Nguyen:2022zwb,
        Baryakhtar:2017dbj
        %	Garani:2018kkd, 
        %	Chen:2018ohx, 
        %	Bell:2018pkk,
	%Acevedo:2019agu, 
	% Garani:2019fpa, %% leptons
	%Acevedo:2019gre, 
	%Hamaguchi:2019oev,
	%Joglekar:2019vzy,
	%Joglekar:2020liw,
	%Bell:2020jou,
	%Dasgupta:2020dik,
	%Garani:2020wge,
	%Bell:2020lmm,
	%Bell:2020obw,
	%Anzuini:2021lnv
	}. 
For example, neutron stars have a high density that can attract and accelerate dark matter particles in the halo to speeds close to $c$. These in-falling dark matter particles can undergo scattering with visible matter in the star, lose kinetic energy, and get gravitationally bound. They can keep losing kinetic energy via repeated collisions and heat up the star. For old, cold neutron stars, the temperature can increase from $\mathcal{O}(100)$~K to $\mathcal{O}(1000)$~K~\cite{Baryakhtar:2017dbj} due to dark matter kinetic heating. Observations of such stars can put strong constraints on dark matter interactions with the Standard Model~\cite{Baryakhtar:2017dbj,
        Raj:2017wrv,
	Garani:2018kkd, 
	Chen:2018ohx, 
	Bell:2018pkk, 
	Acevedo:2019agu, 
	Garani:2019fpa, %% leptons
	Acevedo:2019gre, 
	Hamaguchi:2019oev,
	Bell:2019pyc, %% leptons
        Joglekar:2019vzy,
	Joglekar:2020liw,
	Bell:2020jou,
	Dasgupta:2020dik,
	Garani:2020wge,
        Bell:2020lmm,
	Bell:2020obw,
	Jho:2020sku,
        Anzuini:2021lnv,
        Hamaguchi:2022wpz,
        Fujiwara:2022uiq,
        Chatterjee:2022dhp,
        Coffey:2022eav,
        Acuna:2022ouv}, and these constraints are largely independent of interaction details.

Neutron stars are particularly interesting for the search of inelastic dark matter, see, e.g.,~\cite{Han:1997wn,Tucker-Smith:2001myb, Arkani-Hamed:2008hhe,Chang:2008gd,Cui:2009xq,Schutz:2014nka,DAgnolo:2015ujb,Berlin:2018jbm,Blennow:2016gde,Zhang:2016dck,Alvarez:2019nwt,Bramante:2016rdh,Essig:2018pzq,Tsai:2019buq,Huo:2019yhk,CarrilloGonzalez:2021lxm,Fitzpatrick:2021cij,ONeil:2022szc,Cheng:2022esn}. For a typical inelastic dark matter model, there are two states with a small mass splitting between them. If the universe at present is populated by the light state, terrestrial constraints from direct detection experiments can be significantly weakened. These limits even disappear if the ratio of the splitting to the mass is greater than $10^{-6}$, as there is not enough kinetic energy available in the light state hitting the detector at a speed of $300$ km/s to scatter into a heavy state. However, this is no longer the case when the particle can be accelerated to relativistic speeds. %Since the escape velocity from the surface of a neutron star is about $0.6c$, the in-falling dark matter from the halo is accelerated to relativistic speeds. 
The in-falling dark matter from the halo is accelerated to the escape velocity $\sim 0.6 c$ of a neutron star. This large kinetic energy can allow the light state to overcome the mass gap and facilitate inelastic scattering, leading to observable signals. The kinetic heating due to inelastic dark matter that couples to nucleons has been studied in the effective field theory framework~\cite{Bell:2018pkk}. 

%In particular the kinetic heating of inelastic dark matter that couples to nucleons has been studied previously in the effective field theory framework~\cite{Bell:2018pkk}. 

In this work, we develop a fully relativistic formalism to study neutron star heating with inelastic dark matter. We systematically investigate complex kinematics of the inelastic scattering process between nonrelativistic dark matter particles and ultrarelativistic targets in the star, such as electrons, based on the approach originally developed in the elastic limit~\cite{Joglekar:2020liw}. We then project neutron star constraints on inelastic dark matter interactions with the Standard Model and compare them with those from terrestrial searches including direct detection and collider experiments, see, e.g.,~\cite{Berlin:2018jbm}. Our analysis is for (i) effective vector-vector interactions, (ii) a generic vector mediator, (iii) a benchmark simplified model, where the dark matter candidate is a pseudo-Dirac fermion and it couples to a dark photon. We will demonstrate that compared to the terrestrial searches, our projected constraints from neutron star heating can not only be competitive with existing and future terrestrial experiments in large regions of the parameter space, but also systematically outperform irrespective of interaction details under the consideration. In addition, we will show that neutron stars can be sensitive to inelastic self-interacting dark matter models proposed in~\cite{Blennow:2016gde,Zhang:2016dck,Alvarez:2019nwt}, which are difficult to probe in terrestrial detection. 

The rest of the paper is organized as follows. In Sec.~\ref{sec:2}, we discuss kinematics of neutron star heating with  inelastic dark matter and derive analytical limits on %the phase space of inelastic scattering kinematics. 
maximum allowed mass gap. These formulas provide insight into our numerical results presented in later sections, and they help us estimate constraints on the effective cutoff scale or the coupling constant. In Sec.~\ref{sec:3}, we present the projected constraints assuming vector-vector effective interactions between dark matter and Standard Model particles, as well as a generic vector mediator, and compare them with those from direct detection experiments. We project the neutron star constraints for a concrete pseudo-Dirac inelastic dark matter model with a heavy dark photon mediator in Sec.~\ref{sec:4}, and a light dark photon mediator in Sec.~\ref{sec:5}, where we make comparisons with current and future terrestrial experiments. We conclude in Sec.~\ref{sec:6}. In Appendix~\ref{sec:AppA}, we provide detailed derivation on relativistic inelastic dark matter capture. Appendix~\ref{sec:AppB} shows the derivation of direct detection constraints on inelastic dark matter. 

\section{Kinematics of neutron star inelastic heating}
\label{sec:2}

%Refs.~\cite{Joglekar:2019vzy,Joglekar:2020liw} considered spin-0 and spin-1/2 elastic dark matter candidates and explored the kinematics of neutron star heating for contact interactions with all dimension-5 and -6 effective operators. We extend the study to the capture of inelastic dark matter on relativistic targets as the electrons in the neutron star are ultrarelativistic. We obtain conditions on the maximum mass splitting between the dark matter states, which will help us understand characteristic features exhibited in the numerical results as we will show in the next section.

%Inelastic kinetic heating due to scattering of spin-1/2 heavy dark matter with nucleons via contact interactions was first discussed in~\cite{Bell:2018pkk}. ($\delta m_\text{max}$)

In this section, we present a general formalism for understanding kinematics of inelastic dark matter scattering with relativistic targets in a neutron star, an extension of previous studies for elastic dark matter~\cite{Joglekar:2019vzy,Joglekar:2020liw}, and highlight key differences with respect to the elastic limit. We then obtain the maximum mass splitting between two dark matter states allowed by the capture conditions, which will help us understand characteristic features exhibited in the numerical results as we will show in the next section. More detailed derivations are presented in Appendix~\ref{sec:AppA}. 

We assume (i) a generic inelastic dark matter model consisting of two species $\chi_1$ and $\chi_2$ with masses $m_1$ and $m_2$, respectively, which are separated by a small mass splitting $\delta m\equiv m_2-m_1$; (ii) a vector portal to allow interactions between visible and dark sectors; (iii) off-diagonal portal couplings, thus making inelastic scattering a primary detection mode. We further assume that the dark matter relic density consists of the light species $\chi_1$. 

Under these assumptions, the $\chi_1\,\xi \rightarrow \chi_2\, \xi$ is the predominant scattering mode for dark matter particles to get captured in a neutron star via scattering with Standard Model particles $\xi$ inside the star. Dark matter particles can attain $\gamma_\text{esc}\sim1.25$ at the star's surface due to its strong gravitational field~\cite{Joglekar:2020liw}. If a dark matter particle can lose kinetic energy greater than that in the halo during the transit, then it can get gravitationally bound to the star and said to be captured. If all kinetic energy of the passing dark matter particles can be deposited in the star, its temperature can be heated up to $1600$~K for the stellar radius $R_\star=12.6$~km and mass $M_\star=1.5~{\rm M_\odot}$. More specifically, the temperature depends on the capture efficiency $f$ as~\cite{Baryakhtar:2017dbj, Raj:2017wrv}
\begin{align}
T&=1600\,f^{1/4}\,\text{K}.
\end{align}
If captured dark matter particles can annihilate, the maximum temperature can be as high as $\sim2400$~K for $f=1$~\cite{Raj:2017wrv}. 

Similar to the elastic case~\cite{Joglekar:2019vzy,Joglekar:2020liw}, invariance of total energy in the Center of Momentum (CM) frame ($\sqrt{s}=E_\text{\tiny{CM}}$) allows us to calculate the magnitude of the momentum of dark matter final state in the CM frame $k'_\text{\tiny{CM}}$ as 
\begin{align}
k'^2_\text{\tiny{CM}}&=k^2_\text{\tiny{CM}}-\frac{\left(m_2^2-m_1^2\right)\left(2E^2_\text{\tiny{CM}}+2m^2_\xi-m_1^2-m_2^2\right)}{4E^2_\text{\tiny{CM}}},\label{eqn:kscatcm}
\end{align}
where $k_\text{\tiny{CM}}$ is the magnitude of initial dark matter momentum in the CM frame and $m_\xi$ is the target mass. In the case of inelastic dark matter, the energy transferred to the target in the neutron star frame can be written as
\begin{align}
\Delta E_\text{NS}&=\gamma\left(\sqrt{m_1^2+k_\text{\tiny{CM}}^2}-\sqrt{m_1^2+k'^2_\text{\tiny{CM}}}\right)+\gamma\left(\boldsymbol{\beta}\cdot\bf{k}_\text{\tiny{CM}}\right)\left(1-\frac{k'_\text{\tiny{CM}}}{k_\text{\tiny{CM}}}\cos\psi\right)\notag\\
&\quad-\frac{k'_\text{\tiny{CM}}}{k_\text{\tiny{CM}}}\gamma\sqrt{\beta^2k^2_\text{\tiny{CM}}-\boldsymbol{\beta}\cdot \bf{k}_\text{\tiny{CM}}}\,\sin\psi\cos\alpha,\label{eqn:deltaE}
\end{align}
where $\psi$ and $\alpha$ are the polar and azimuthal angles of scattering in the CM frame, respectively. See Appendix~\ref{sec:AppA} for descriptions of CM frame quantities in terms of the momenta, energies and angles in the neutron star frame and other details. We discuss key differences with respect to the elastic case in what follows.

\begin{table}[t]
	\centering
	\begin{tabular}{p{1.5cm}p{2.1cm}p{2.3cm}p{1.8cm}p{2.3cm}p{2.3cm}p{2.3cm}}  
		\toprule
		$m_\xi$
		& 
		\multicolumn{2}{l}{Nonrelativistic}
		& 
		\multicolumn{3}{l}{Relativistic}
		\\
		\cmidrule(r){1-3}\cmidrule(r){4-7}
		$m_1$
		& % NRH 
		{\small Heavy}
		& % NRL
		{\small Light}
		& % RH
		{\small Heavy}
		& % RL
		{\small Light-ish}
		&
		{\small Med.~Light}
		% M.~Light
		& % RVL
		% V.~Light
		{\small Very~Light}
		\\
		\midrule
		$\delta m_\text{max}$
		&
		$(\gamma_\text{esc}-1)m_\xi$
		&
		$(\gamma_\text{esc}-1)m_1$
		&
		$2\gamma_\text{esc}\beta_\text{esc}p^\text{F}_\xi$
		&
		$\frac{\left(\gamma_{\rm esc}^2-1\right)}{2}m_1$
		&
		$\frac{\left(\gamma_{\rm esc}^2-1\right)}{2}m_1$
		&
		$\frac{\left(\gamma_{\rm esc}^2-1\right)}{2}m_1$
		\\
		\bottomrule
	\end{tabular}
	\caption{The maximum mass splitting between two dark matter states, above which the upscattering process $\chi_1\xi\rightarrow\chi_2\xi$ is kinematically forbidden in a neutron star. %The values of $\delta_\text{max}$ as per Eq.~\ref{eqn:conds} up to the lowest order. The values obtained in nonrelativistic target case match $\delta m_\text{max}$ evaluated in previous study of iDM capture in neutron stars with nonrelativistic targets~\cite{Bell:2018pkk}.
		}
	\label{tab:dmax}
\end{table}

%It is important to note where inelastic scattering departs from elastic scattering. 

%We first note that for both non-relativistic and relativistic targets, the momentum and energy transferred is saturated for $m_1>\text{Max}[m_\xi,p^\text{F}_\xi]\sim 1$~GeV~\cite{Joglekar:2019vzy}. 

For inelastic scattering $\chi_1\,\xi\rightarrow\chi_2\,\xi$ to occur, the target $\xi$ must be knocked out of its Fermi surface, and hence there is an upper limit on the percentage of initial kinetic energy available for transition from $m_1$ to $m_2$. Since maximum kinetic energy available in incoming dark matter is $\sim \left(\gamma_\text{esc}-1\right)\,m_1$%at the surface of a neutron star (and about $10-20\%$ more if it can penentrate deep inside without scattering)
, the maximum value of the relative mass gap $\Delta\equiv(m_2-m_1)/m_1=\delta m/m_1$ is $\left(\gamma_\text{esc}-1\right)$. This can be achieved for $m_1<1$~GeV. For heavy dark matter masses $m_1>1$~GeV, the energy transferred is saturated and proportional to the target mass $m_\xi$ and the Fermi momentum $p^\text{F}_\xi$ for nonrelativistic and ultrarelativistic targets~\cite{Joglekar:2019vzy}, respectively (as implied by the maximum mass gaps shown in Table~\ref{tab:dmax}). Thus the maximum allowed $\Delta$ value scales as $m_1^{-1}$. This feature leads to a sharp drop-off in the sensitivity of neutron star heating for large dark matter masses when $\Delta$ is fixed, i.e., $m_1\gtrsim m_\xi/\Delta$, $p^{\rm F}_\xi/\Delta$ for nonrelativistic and ultrarelativistic targets, respectively. Conversely, if the mass gap $\delta m$ is fixed, then heavy dark matter will always get captured as long as $\delta m$ is less than the maximum mass gap $\delta m_\text{max}$. %, as summarized in Table~\ref{tab:dmax}. 

Conditions on $k'_\text{\tiny{CM}}$ and $\Delta E_\text{\tiny{NS}}$ will determine the maximum mass gap $\delta m_\text{max}$ that can lead to successful capture for a given dark matter mass. These conditions follow from the facts that $k'_\text{\tiny{CM}}$ should have a real value that is also less than $k_\text{\tiny{CM}}$, and $\Delta E_\text{\tiny{NS}}$ should be large enough to knock the target out of its Fermi surface, i.e.,
\begin{align}
0<k'^2_\text{\tiny{CM}}&<k^2_\text{\tiny{CM}};\quad \Delta E_\text{\tiny{NS}}+E_\xi>E_\xi^\text{F}.\label{eqn:conds}
\end{align}
Using Eqs.~\ref{eqn:kscatcm} and~\ref{eqn:deltaE} and retaining the lowest order terms, we obtain $\delta m_\text{max}$ for different cases as summarized in Table~\ref{tab:dmax}; see Appendix~\ref{sec:AppA} for details. For nonrelativistic targets, supplying enough energy to the dark matter particle for overcoming the mass gap is the limiting factor. Thus, the first condition of Eq.~\ref{eqn:conds} is most restrictive. For relativistic targets, the main constraint arises from the ability of dark matter to knock the target out of its Fermi surface, making the second condition of Eq.~\ref{eqn:conds} most relevant.

%We have discussed the role of these two conditions of Eq.~\ref{eqn:conds} in determining the maximum mass gaps tabulated in Table~\ref{tab:dmax} for which the capture is still permitted, in Appendix~\ref{App:B}. 

We demonstrate the derivation of the maximum mass gap to the first order in the case of very light dark matter ($m_1\ll m_\xi^2/p^\text{\tiny{F}}_\xi$). In this case, we have $x\ll z^2\ll z \ll 1$,
where $x=m_1/p_\text{\tiny{F}}$, $y=\delta m/p_\text{\tiny{F}}$ and $z=m_\xi/p_\text{\tiny{F}}$.\footnote{In the calculations that follow, the target type $\xi$ is fixed. Hence, to shorten the notation in the rest of the paper, we drop the subscript $\xi$ for momenta to write $p$ and $p_\text{\tiny{F}}$ instead of $p_\xi$ and $p^\text{\tiny{F}}_\xi$.} We take $\Delta E_\text{\tiny{NS}}+E_\xi-E^\text{\tiny{F}}_\xi$, and expand in $y$, $x$ and $z$ sequentially. As shown in Appendix~\ref{sec:AppA}, the optimum CM frame scattering angle $\psi$ for maximum energy transfer are given by
\begin{align}
1-\cos\psi<\beta_\text{esc}^2z^2/2;\quad \sin\psi<\beta_\text{esc}z.
\label{eqn:angle:max2}
\end{align} 
Thus, $(1-\cos\psi)$ is of the order $z^2$ and $\sin\psi$ is of the order $z$. Corresponding optimum choices for maximum energy transfer to the star are $\sin\theta=1$ and $\cos\alpha=-1$, where $\theta$ is the angle between the dark matter and the target momenta in the neutron star frame. Using these optimum values for maximum energy transfer to the targets, we write
\begin{align}
E_\xi-E^\text{\tiny{F}}_\xi+\Delta E_\text{\tiny{NS}}&\approx p-p_\text{\tiny{F}}-\frac{\beta_\text{esc}^2\gamma_\text{esc}}{2}\frac{p\,^2}{p_\text{\tiny{F}}}x+\beta^2_\text{esc}\gamma_\text{esc}\,p\,x-\frac{p_\text{\tiny{F}}}{\gamma_\text{esc}}y.
\label{eqn:verylight}
\end{align}

%The terms on the right hand side of Eq.~\ref{eqn:verylight} all appear to be combined first orders in $x$, $y$ and $z$, but some are actually higher-order terms, because upper bounds for the expressions like $(1-\cos\psi)$ or $\sin\psi$ have powers of $x$, $y$ and $z$ hidden in them. 

We have retained only first order terms in Eq.~\ref{eqn:verylight}.
%we find 
%\begin{align}
%\Delta E_\text{\tiny{NS}}+E_\xi-E_\xi^\text{F}&\approx p-p_\text{\tiny{F}}-\frac{(1-\cos\psi)\gamma_\text{esc}p\,^2 (1-\beta_\text{esc}\cos\theta)}{p_\text{\tiny{F}}}\frac{x}{z^2}+(-\cos\alpha)\beta_\text{esc}\gamma_\text{esc}\,(\sin\theta)\sin\psi\,p\frac{x}{z}\notag\\
%&\quad\quad -\frac{(\cos\psi)\, p_\text{\tiny{F}}}{\gamma_\text{esc}(1-\beta_\text{esc}\cos\theta)}y.	
%\end{align}
We can use the second condition in Eq.~\ref{eqn:conds} for targets on the edge of the Fermi surface with maximum possible scattering angle compatible with Eq.~\ref{eqn:angle:max2} to get an upper bound on $\delta m$,
\begin{align}
%y&<\frac{1}{2}\beta_\text{esc}^2\gamma_\text{esc}^2\,x\quad\longrightarrow\quad 
\delta m<\frac{1}{2}\,\beta_\text{esc}^2\gamma_\text{esc}^2\,m_1.
\end{align}
More details of the derivation for different cases of targets and dark matter mass ranges can be found in Appendix~\ref{sec:AppA}.

Before proceeding further, we comment on the thermalization process of dark matter particles with Standard Model particles in a neutron star. The $\gamma$ factor of a dark matter particle reduces as it loses energy. Therefore, the maximum mass gap that can be overcome in each of the successive collisions will reduce. Consider the case with nonrelativistic targets and heavy dark matter. Using the results in Table~\ref{tab:dmax}, we get a lower bound on the kinetic energy to be $(\gamma-1)m_1\approx m_1^2\Delta/m_\xi$ that the dark matter must posses in order to undergo subsequent scatterings. Dark matter trajectories that pass deeper inside the star will have more kinetic energy available to be transferred, because $\gamma_\text{esc}$ increases up to $20\%$ deeper inside the star. %As a result, larger percentage of kinetic energy can be transferred to the star. 
One can estimate the percentage of total kinetic energy transferred by dark matter particles at various depths in the neutron star core, and the weighted average of these percentages can give the approximate maximum percentage of total kinetic energy transferable by all dark matter particles to the star. We estimate that for an example value of $m_1\Delta/m_\xi\sim 0.1$, about $70\textup{--}75\%$ of the kinetic energy can be transferred, thus resulting in the temperatures of $90\textup{--}95\%$ of what is achievable in the elastic case. 

For ultrarelativistc targets and heavy dark matter, the lower bound on kinetic energy is $(\sqrt{1+{m_1^2\Delta^2}/{4p^2_\text{F}}}-1)m_1$. A rough estimate shows about $70\textup{--}75\%$ of kinetic energy transfer for $m_1\Delta/2p_\text{F}\sim 0.45$ as an example. In the case of light dark matter, the Pauli blocking is the limiting condition, where a large percentage of interactions occur if the particle transfers kinetic energy $\mathcal{O}(0.1)$ times its original energy. This will naturally help transfer almost all the kinetic energy, if the interaction is allowed at all by Pauli blocking. %This is reflected in Table~\ref{tab:dmax} as well, where 
The lower bound on the kinetic energy that the dark matter particle must possess for the inelastic interaction is independent of the target mass or momentum. It is simply given as $m_1\Delta$ for nonrelativistic targets and $(\sqrt{2\Delta+1}-1)m_1$ for ultrarelativistic targets. These relations can be inferred from Table~\ref{tab:dmax} as well.

For simplicity, we will show regions of parameter space assuming $f=1$ for the rest of the paper.  The subtlety associated with thermalization process implies that in the region with $f=1$, the attainable temperature from kinetic heating of inelastic dark matter is greater than $90\textup{--}95\%$ of that expected in the elastic case, when the conditions $m_1\Delta\le 0.1 m_\xi$ and $0.9 p_\text{F}$ are satisfied for nonrelativistic and ultrarelativistic targets, respectively. For higher $m_1$ towards the boundary of $f=1$ region, at which a sharp drop-off occurs, the maximum attainable temperature of the star reduces below $\sim1500$~K for non-annihilating dark matter. Therefore, the actual $m_1$ reach from neutron star heating could reduce by an $\mathcal{O}(1)$ factor in the non-annihilating case. As we will show, for the most parameter space that we are interested, the constraints from neutron stars can significantly surpass those from terrestrial experiments, and the relative comparison does not change even after taking into account the reduction effect due to thermalization. 

\section{Projected constraints on inelastic dark matter with a generic interaction}
\label{sec:3}

In the optically thin limit, the capture probability for a dark matter particle transiting through a neutron star can be calculated as~\cite{Joglekar:2019vzy}
\small
%\begin{widetext}
\scriptsize
    \begin{equation}
	{f}=
	\sum
	\limits_{N_{\rm hit}\,\in\,\mathbb{Z}}
	\frac{\left<n_{\rm T}\right>\Delta t}{N_{\rm hit}}
	\int d\Omega_\text{NS}
	\int\limits^{p_\text{F}}_0 d|\bar{p}|\frac{|\bar{p}|^2}{V_\text{F}} v_\text{M\o l}
	\int d\Omega_\text{CM}\left(\frac{d\sigma}{d\Omega}\right)_\text{CM}\Theta\left(\Delta E+E_\xi-E_\xi^\text{F}\right)\Theta\left(\frac{E_{\rm halo}}{N_{\rm hit}-1}-\Delta E\right)\Theta\left(\Delta E - \frac{E_{\rm halo}}{N_{\rm hit}} \right).
	\label{eqn:fullf}
	\end{equation}
%\end{widetext}
\normalsize
We first consider two simplified models that describe dark matter interactions with the Standard Model: (i) an effective vector-vector operator between a fermionic dark matter particle and a Standard Model fermion for fixed relative mass gap $\Delta$; (ii) a vector mediator of masses $10$~MeV and $10$~GeV for fixed mass gap $\delta m$. We compare the projected constraints from neutron stars with those from direct detection experiments. For this analysis, we consider electron and proton targets. They are picked to demonstrate the projected constraints of neutron star heating in leptophilic and leptophobic cases. If the dark matter-neutron interaction is present at same level as the dark matter-proton interaction, the constraints from the neutron targets will be stronger as the number of neutrons is much higher. This justifies the use of proton targets to represent conservative constraints in the leptophobic case.  In our numerical study, we take a typical model for neutron stars with $M_\star=1.5\,M_\odot$ and $R_\star=12.6$~km. We only consider the capture in the core with a radius of $11.6$~km conservatively and we assume an average density and $p_\text{\tiny F}$ same as that in~\cite{Joglekar:2020liw} as per BSk-24 equation of state~\cite{Pearson:2018tkr}. 

\begin{figure}[tp]
	\centering
	\includegraphics[width=.48\textwidth]{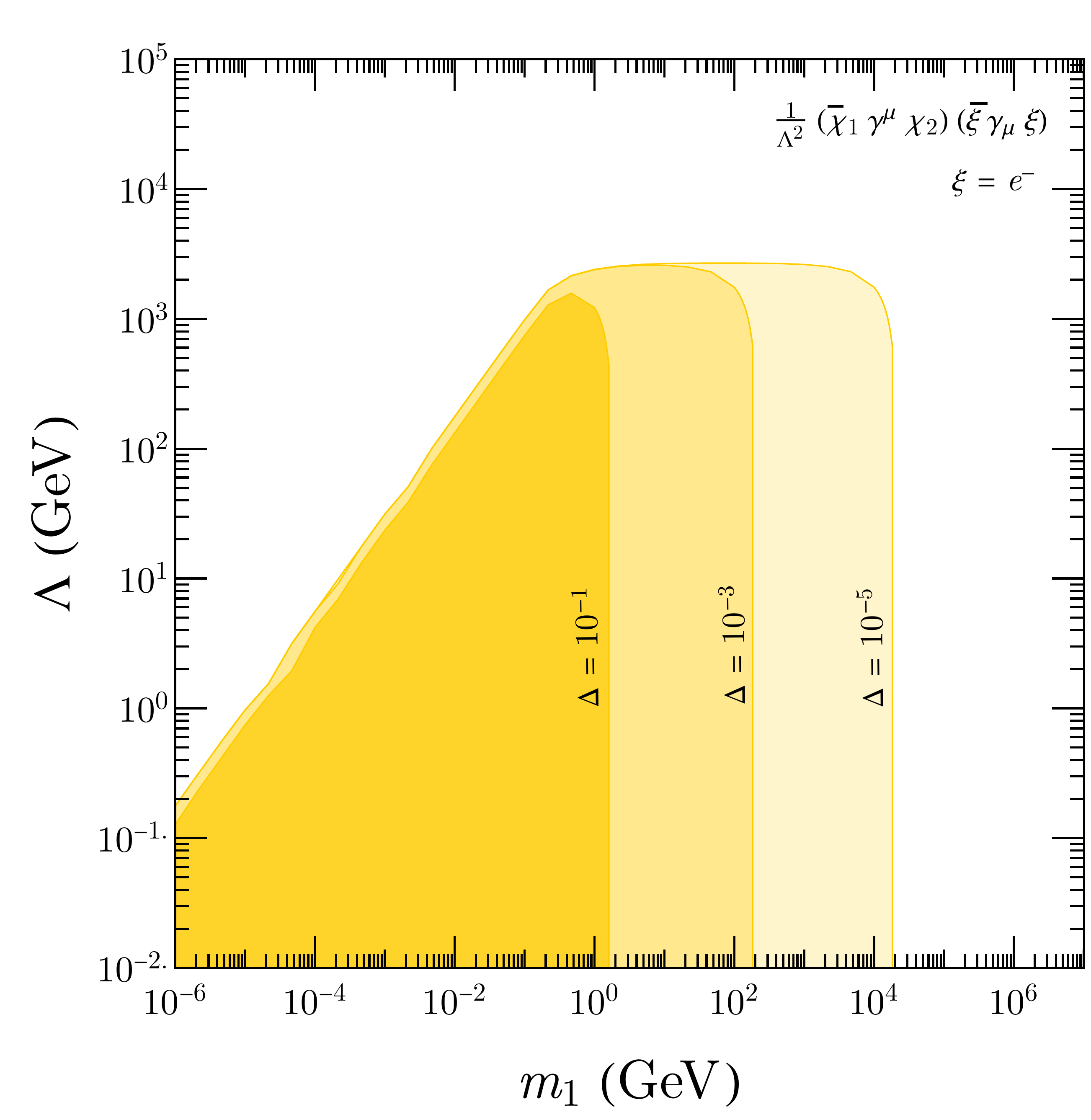}
	\includegraphics[width=.48\textwidth]{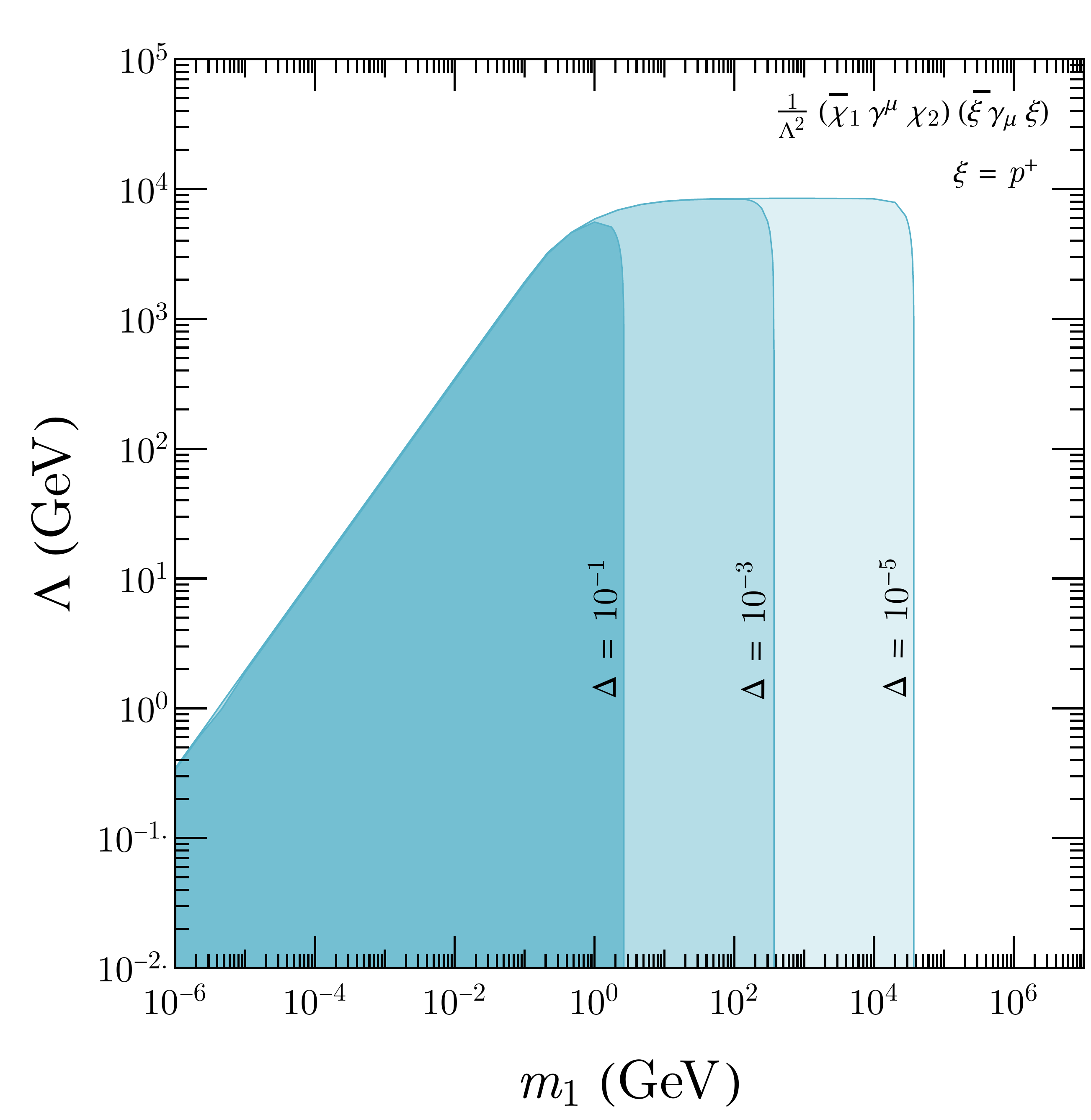}
	\caption{Projected constraints on the cutoff scale $\Lambda$ from neutron stars, assuming effective contact interactions with electron (yellow-shaded, left panel) and proton (cyan-shaded, right panel) targets. The parameter $\Lambda$ is the cutoff scale for the contact operator. The relative mass splittings are $\Delta=(m_2-m_1)/m_1=10^{-1},~10^{-3}$ and $10^{-5}$. }
	\label{fig:contact} 
\end{figure}

In Fig.~\ref{fig:contact}, we present constraints on the cutoff scale $\Lambda$ for the effective operator from neutron star heating for electron (left panel) and for proton (right panel) targets. We take $\Delta = 10^{-1}$, $10^{-3}$ and $10^{-5}$ that are inaccessible in direct detection experiments. For low dark matter masses, the projected constraints are limited by Pauli blocking. As the mass becomes larger, the reach eventually saturates as $m_1\sim m_\xi$ and $p_{\text {\tiny F}}$ for proton and electron targets, respectively, similar to the elastic case~\cite{Joglekar:2019vzy,Joglekar:2020liw}. A sharp drop-off can be clearly seen in Fig.~\ref{fig:contact} for large masses corresponding to a maximum allowed relative mass gap.
Consider electron targets and heavy dark matter, $\delta m_{\rm max}\approx2\gamma_{\rm esc}\beta_{\rm esc}p_{\tiny{\rm F}}$ as in Table~\ref{tab:dmax}. Thus, the drop-off should occur when $m_1\approx\delta m_{\rm max}/\Delta\approx218$~GeV,  
where we have taken $\beta_{\rm esc}=0.6$, $\gamma_{\rm esc}=1.25$, $p_{\tiny{\rm F}}=145$~MeV, and $\Delta=10^{-3}$. This estimate is consistent with the numerical calculation. The results in the right panel of Fig.~\ref{fig:contact} are similar to the ones in~\cite{Bell:2018pkk} obtained for the neutron target, after taking into account the difference in proton and neutron abundances in neutron stars. Note the vertical drop-off in $m_1$ can be lowered by a small $\mathcal{O}(1)$ factor due to the incompletion of thermalization, as discussed in the last section.
 %These values are within $\mathcal{O}(10\%)$ of the leading order analytical estimates derived in Table~\ref{tab:dmax}. Since the figure is plotted for fixed \textit{relative} mass gap, larger dark matter masses $m_1$ mean larger mass splitting. If the mass splitting reaches the maximum allowed mass gap then the up-scattering is not allowed kinematically and the there is no NS heating, which leads to the sharp vertical cut-off. 

\begin{figure}[tp]
	\centering
	\includegraphics[width=.48\textwidth]{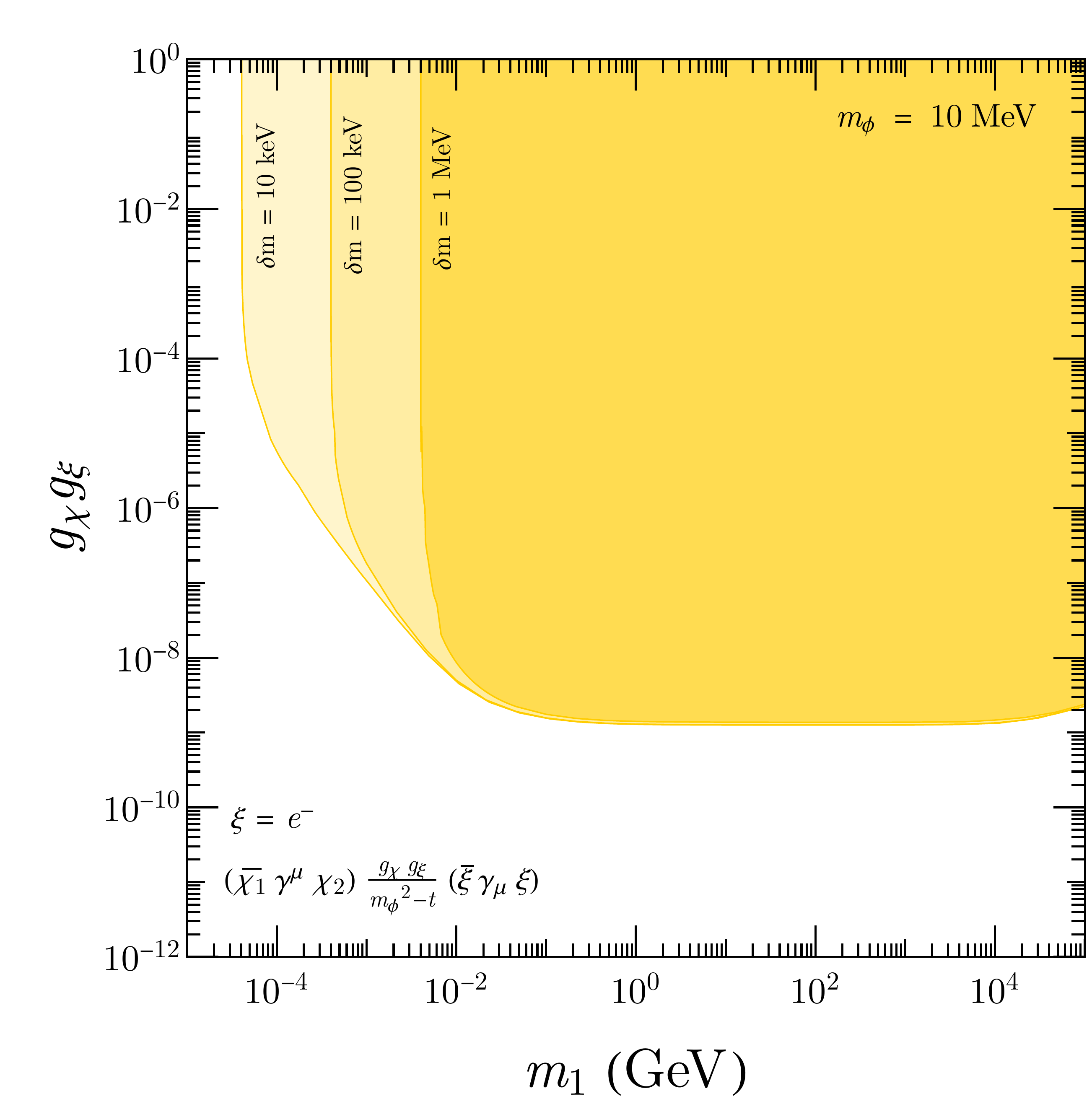}
	\includegraphics[width=.48\textwidth]{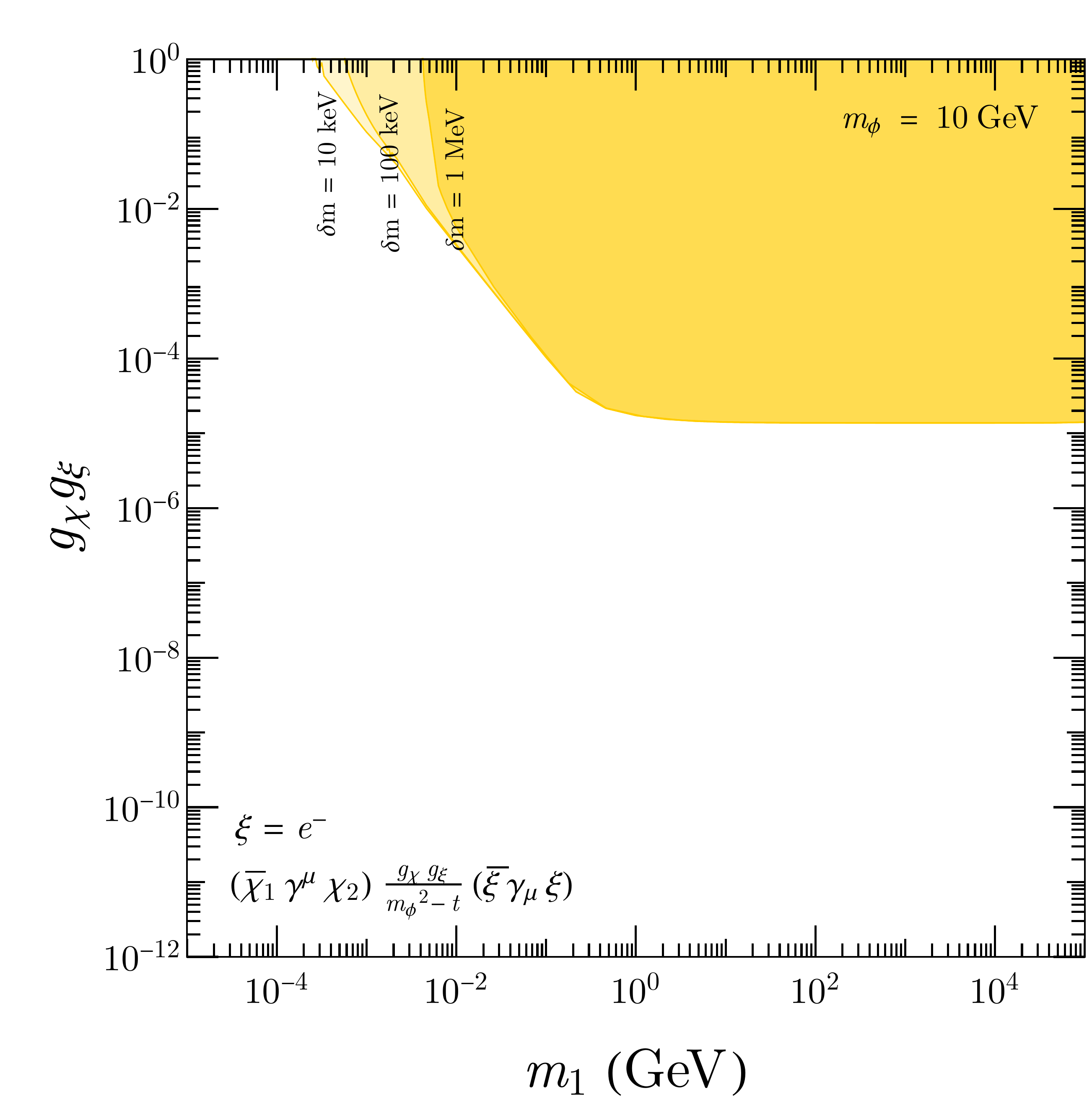}\\
	\includegraphics[width=.48\textwidth]{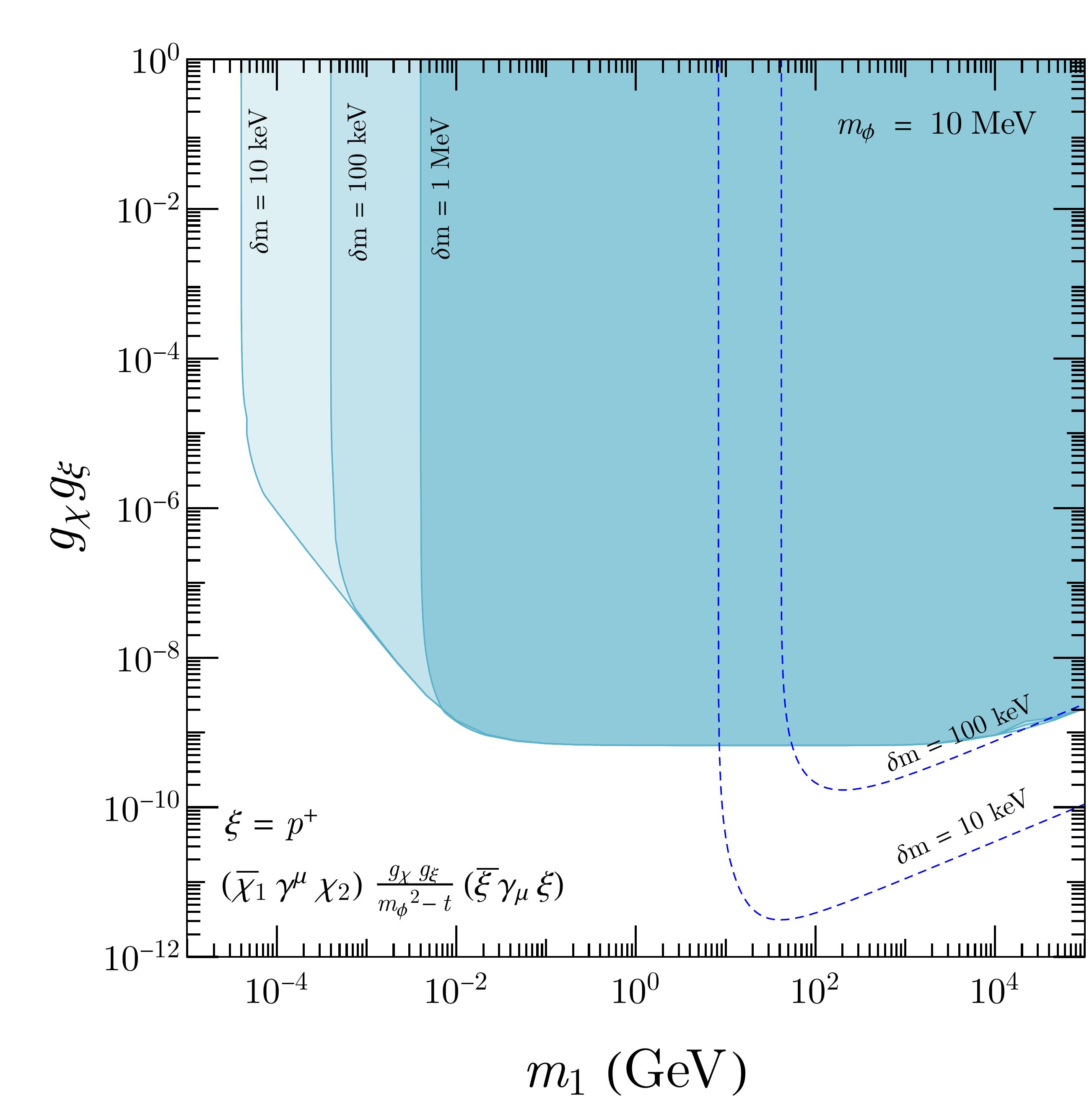}
	\includegraphics[width=.48\textwidth]{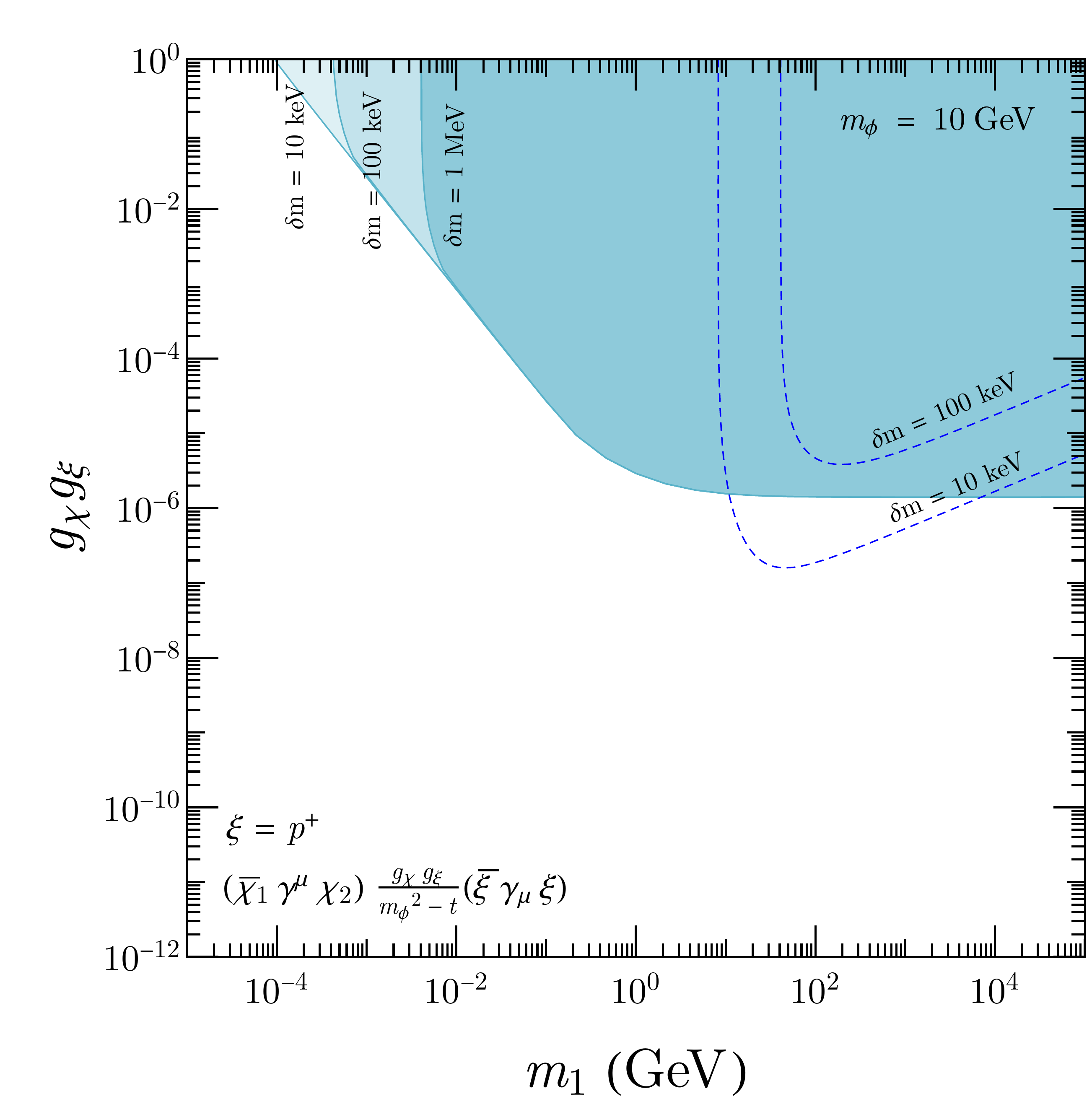}
	\caption{
	Projected constraints on the combination of coupling constants $g_\chi g_\xi$ from neutron stars, assuming portal interactions with electron (yellow-shaded, top panels) and proton (cyan-shaded, bottom panels) targets for fixed mass gaps $\delta m=10,~100$, and $1000$ keV. The mediator masses are taken to be $m_\phi=10$ MeV (left panels) and $10$ GeV (right panels). Regions above the dashed-blue curves are excluded in terrestrial direct detection experiments for the mass gap indicated on individual contours.}
	\label{fig:fixedgap} 
\end{figure}

We also consider constraints for the model with a vector mediator for fixed mass gap $\delta m$. Higher the mass gap, the event rate in direct detection experiments is more suppressed. For both xenon and germanium detectors, the rate of event in the detector rapidly gets suppressed at a level of $90\textup{--}99.9\%$ for the mass gap $\delta m$ between $10$~keV to $1$~MeV~\cite{Blennow:2016gde}. Therefore, it is instructive to project the constraints from neutron stars in this $\delta m$ range and compare them with the corresponding direct detection bounds. In Fig.~\ref{fig:fixedgap}, we show the projected constraints on the combination of the coupling constants $g_\chi g_\xi$ for mass gaps $\delta m = 10,\,100$, and $1000$~keV. The top panels show the results for electron targets with a generic vector mediator $\phi$ of mass $m_{\phi}=10\,{\rm MeV}$ (left panel) and $m_{\phi}=10\,{\rm GeV}$ (right panel). The bottom panels show those for proton targets. The drop-off is present in the low mass region rather than the high mass region as the $\delta m$ is fixed in this case unlike the fixed $\Delta$ case, where $\delta m$ rises with $m_1$. The drop-off values in $m_1$ from our numerical calculation agree with those using analytical expressions in Table~\ref{tab:dmax} within $\sim10\%$. Consider the case of proton targets and light dark matter as shown in the bottom two panels of Fig.~\ref{fig:fixedgap}. We know $\delta m_{\rm max}\approx(\gamma_{\rm esc}-1)m_1$ from Table~\ref{tab:dmax}. Hence, inelastic scattering does not occur if $m_1\lesssim\delta m/(\gamma_{\rm esc}-1)=400$~keV for $\delta m=100$~keV and $\gamma_{\rm esc}=1.25$.

The kinetic heating constraints have two important features in the case of fixed $\delta m$, irrespective of the target particles. For heavy dark matter, $\delta m_{\rm max}$ is independent of $m_1$, and thus the projected constraints are insensitive to $m_1$ until multi-scattering becomes relevant for $m_1>1$~PeV. In contrast, the direct detection limits start weakening for $m_1\gtrsim100$~GeV, as indicated in the bottom panels of Fig.~\ref{fig:fixedgap} (dashed blue). In addition, the lower threshold on $m_1$ is less than $10\,\delta m$, which is far lower than any current or future direct detection experiment can achieve. There are no direct detection constraints for $\delta m\gtrsim 10$~keV and $10$~MeV for leptophilic and non-leptophilic cases, respectively. For $\delta m=10\textup{--}100$~keV, direct detection constraints, recasted from the XENON1T results~\cite{XENON:2018voc}, can be stronger or comparable to those from neutron stars only when $m_1\gtrsim10\textup{--}100$~GeV; see the bottom panels of Fig.~\ref{fig:fixedgap} and Appendix~\ref{sec:AppB} for details. Thus, neutron stars have potential to complement and exceed direct detection experiments in the search for inelastic dark matter.

\section{Constraining inelastic dark matter with a heavy mediator}
\label{sec:4}

In this section, we compare the projected constraints from neutron stars with those from terrestrial searches for a specific inelastic dark matter model similar to the one discussed in~\cite{Berlin:2018jbm}. The model consists of a psuedo-Dirac fermionic dark matter particle with a light state $\chi_1$ with mass $m_1$ and a heavy state $\chi_2$ of mass $m_2=m_1+\delta m$. The interactions in the dark sector are off-diagonal and mediated by a dark $U(1)$ gauge boson $A'_\mu$. The model assumes a vector portal via the kinetic mixing scenario between dark gauge boson and Standard Model photon. In the mass basis, the diagonalized Lagrangian for the relevant interactions is
\begin{equation}
\mathcal{L}_{\rm int}\supset \dfrac{1}{2}m_{A'}^2 A^{'\mu} A'_{\mu}
-\left(\frac{1}{2}g_{\chi}\bar{\chi}_2\gamma^{\mu}\chi_1A'_{\mu}+h.c.\right)+q\epsilon \bar{\zeta}\gamma^{\mu}\zeta A'_{\mu},
\label{eq:model}
\end{equation} 
where $m_{A'}$ is the dark photon mass, $g_\chi$ is the dark gauge coupling constant, $\epsilon$ is the kinetic mixing parameter and $\zeta$ denotes fermions in the Standard Model with electric charge $q$. We assume that the light state is the dominant component of the dark matter abundance, but see~\cite{Bramante:2020zos, Dutta:2021wbn} for models wheres the heavy state $\chi_{2}$ has a significant abundance. In addition, in this section we focus on $m_{A'}>m_1$, where the freeze-out proceeds through a co-annihilation process to the Standard Model particles $\chi_1\chi_2\rightarrow\zeta\bar{\zeta}$. At late times, the $\chi_2$ abundance is exponentially suppressed. Hence, the co-annihilation rate is negligible, avoiding the indirect detection constraints. We project the neutron star constraints for this parameter space, which is cosmologically viable, and compare them with terrestrial experiments. In the next section, we will consider the case of $m_{A'}<m_1$. 

%If the dark photon is lighter than the dark matter particle $m_{A'}<m_1$, the annihilation to mediator $\chi_1\bar{\chi}_1\rightarrow A'_\mu A'_\mu$ is the dominant mode for setting the $\chi_1$ relic abundance in the early universe. The annihilation is unsuppressed at late times and could face stringent constraints from measurements of cosmic microwave background radiation~\cite{Planck:2018vyg,Bringmann:2016din}. These constraints could be relaxed and alleviated with additional model building work; see~\cite{Bramante:2020zos}. For instance, if there exists a massless fermion in the dark sector to which the mediator predominately decays~\cite{Huo:2017vef}, the branching ratio to the Standard Model particles can be suppressed. We will discuss this case in the next section in the context of secluded and self-interacting thermal dark matter. On the other hand, for $m_{A'}>m_1$, the freeze-out proceeds through a co-annihilation process to the Standard Model particles $\chi_1\chi_2\rightarrow\psi\bar{\psi}$. At late times, the $\chi_2$ abundance is exponentially suppressed and hence the co-annihilation rate is negligible, avoiding the indirect detection constraints. We project the neutron star constraints for this parameter space, which is cosmologically viable, and compare them with terrestrial experiments.

\begin{figure}[tp]
	\centering
	\includegraphics[width=.48\textwidth]{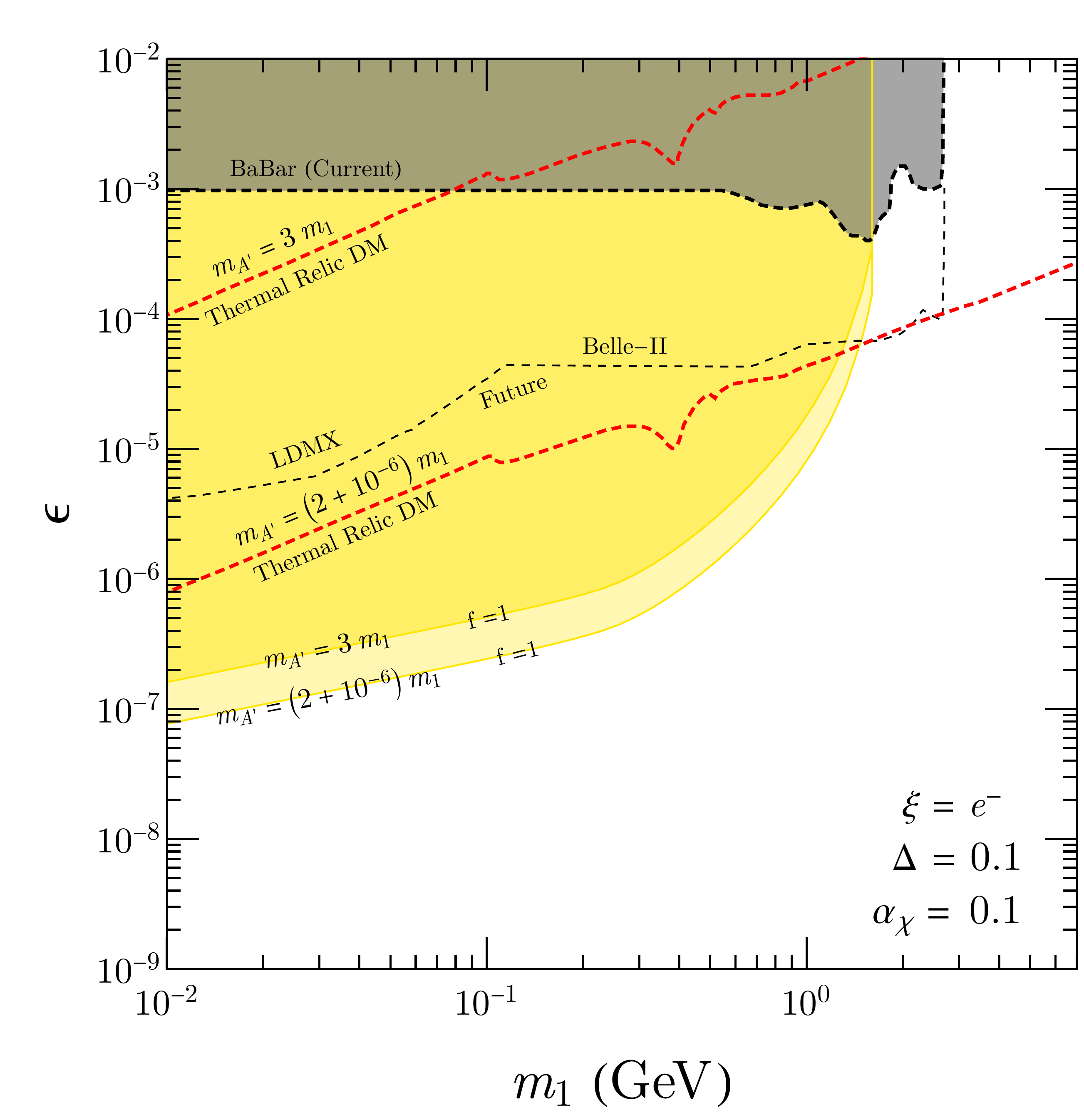}
	\includegraphics[width=.48\textwidth]{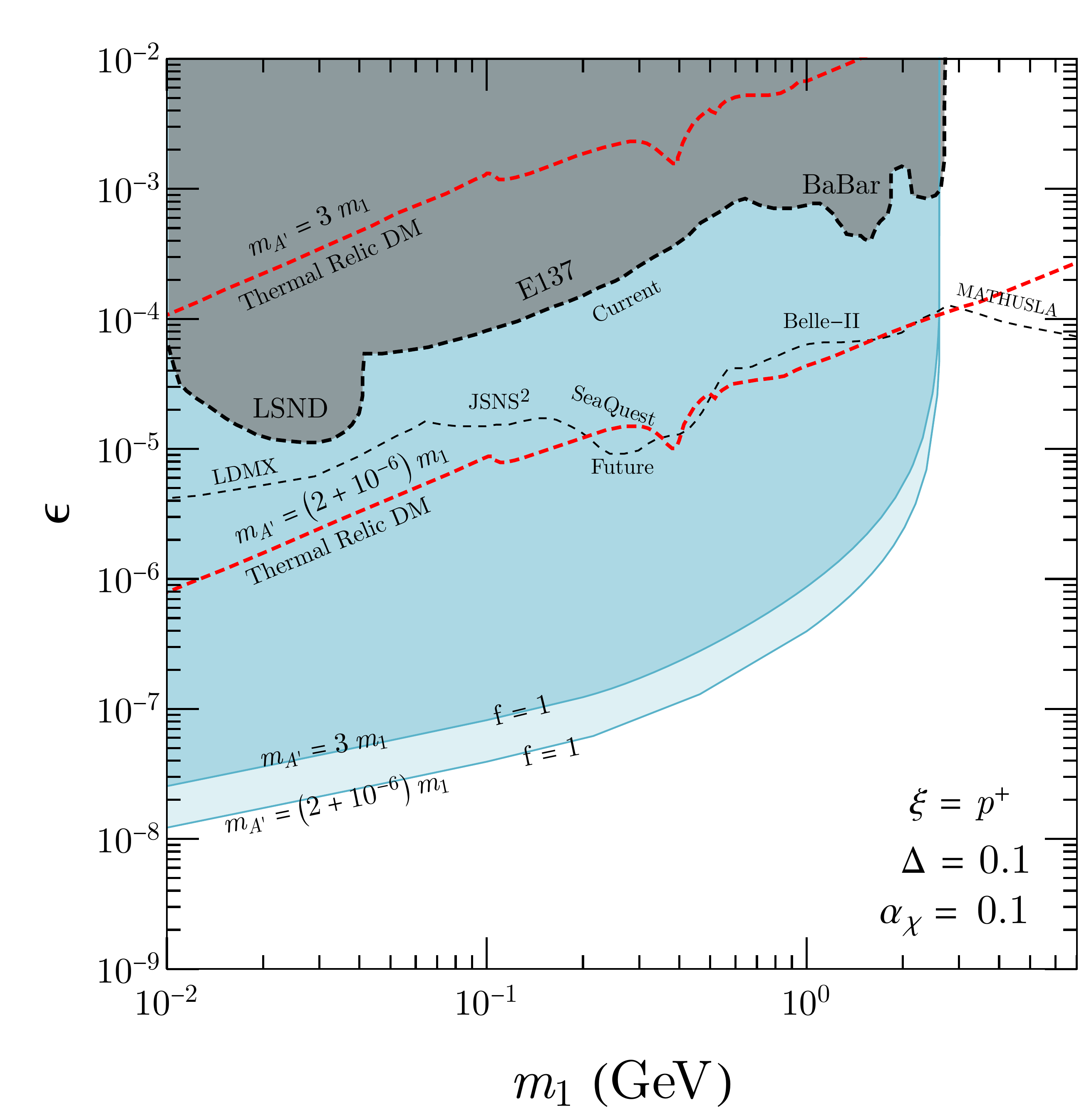}
	\caption{Projected constraints on the kinetic mixing parameter $\epsilon$ from neutron star heating for inelastic dark matter, assuming portal interactions with electron (yellow-shaded, left panel) and proton (cyan-shaded, right panel) targets, where $\Delta=0.1$ and $\alpha_\chi=0.1$. The dark photon masses are $m_{A'}=3m_1$ and $m_{A'}=(2+10^{-6})m_1$. The gray-shaded region is excluded, taken from~\cite{Berlin:2018jbm}, which depends on the bounds from BaBar~\cite{BaBar:2017tiz} (in both panels), E137~\cite{Bjorken:1988as,Batell:2014mga} and LSND~\cite{LSND:2001akn,deNiverville:2011it} (right panel). Regions above the dashed-black curves could be excluded in the future by LDMX~\cite{Berlin:2018bsc,Berlin:2018pwi,LDMX:2018cma} and Belle-II~\cite{Belle-II:2018jsg} (both panels), JSNS$^2$~\cite{Jordan:2018gcd}, SeaQuest~\cite{Berlin:2018pwi}, FASER~\cite{Feng:2017uoz}, MATHUSLA~\cite{Chou:2016lxi}, CODEX-b~\cite{Gligorov:2017nwh}, LHCb~\cite{LHCb:2017trq,Ilten:2016tkc,Pierce:2017taw}, BDX~\cite{Izaguirre:2017bqb}, MiniBoone~\cite{MiniBooNE:2017nqe,MiniBooNEDM:2018cxm} (right panel) as compiled in~\cite{Berlin:2018jbm}. The names of the experiments shown are only the ones that can provide the strongest bounds for a given dark matter mass in order to reduce the clutter. The dotted-red curves denote dark matter as a thermal relic with the dark photon mass indicated.}
	\label{fig:heavymed} 
\end{figure}

%~\cite{Planck:2018vyg}

We further specify the ratio of $m_{A'}$ to $m_1$. Ref.~\cite{Berlin:2018jbm} studied and compiled constraints on the model from current and future accelerator and beam dump experiments assuming $m_{ A'}/m_1=3$. Increasing the $m_{ A'}/m_1$ ratio, while keeping the correct relic abundance, would push the model towards the exclusion region, making it cosmologically less interesting. Thus, we will take $m_{A'}/m_1=3$ for a case study. Furthermore, we consider $m_{A'}/m_1=2+10^{-6}$ as well. In this case, the co-annihilation during freeze-out is enhanced by the resonant effect, the $\epsilon$ value, required to obtain the correct relic density, becomes much smaller~\cite{Feng:2017drg}. The model is less constrained by the terrestrial searches. However, as we will show, the projected constraints from neutron star heating remain relatively \textit{insensitive} to the change of $m_{A'}/m_1$ ratio. Thus, neutron stars can be excellent targets for probing $1<m_{A'}/m_1<3$ region of the parameter space. In order to present and compare our results with the terrestrial searches, we fix $\alpha_\chi=g^2_\chi/4\pi=0.1$ as in~\cite{Berlin:2018jbm}. 
	
\begin{figure}[tp]
	\centering
	\includegraphics[width=.48\textwidth]{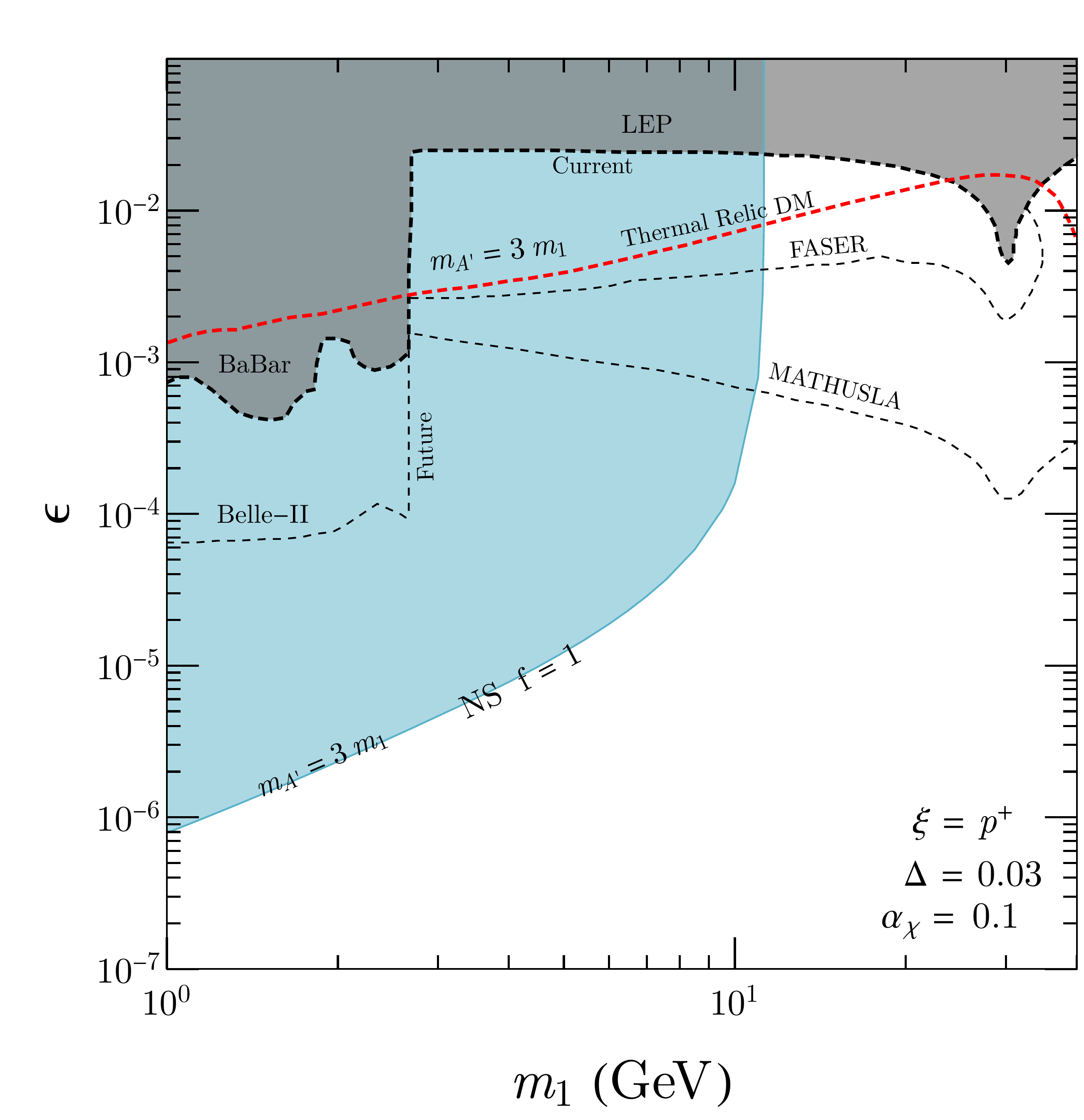}
	\includegraphics[width=.48\textwidth]{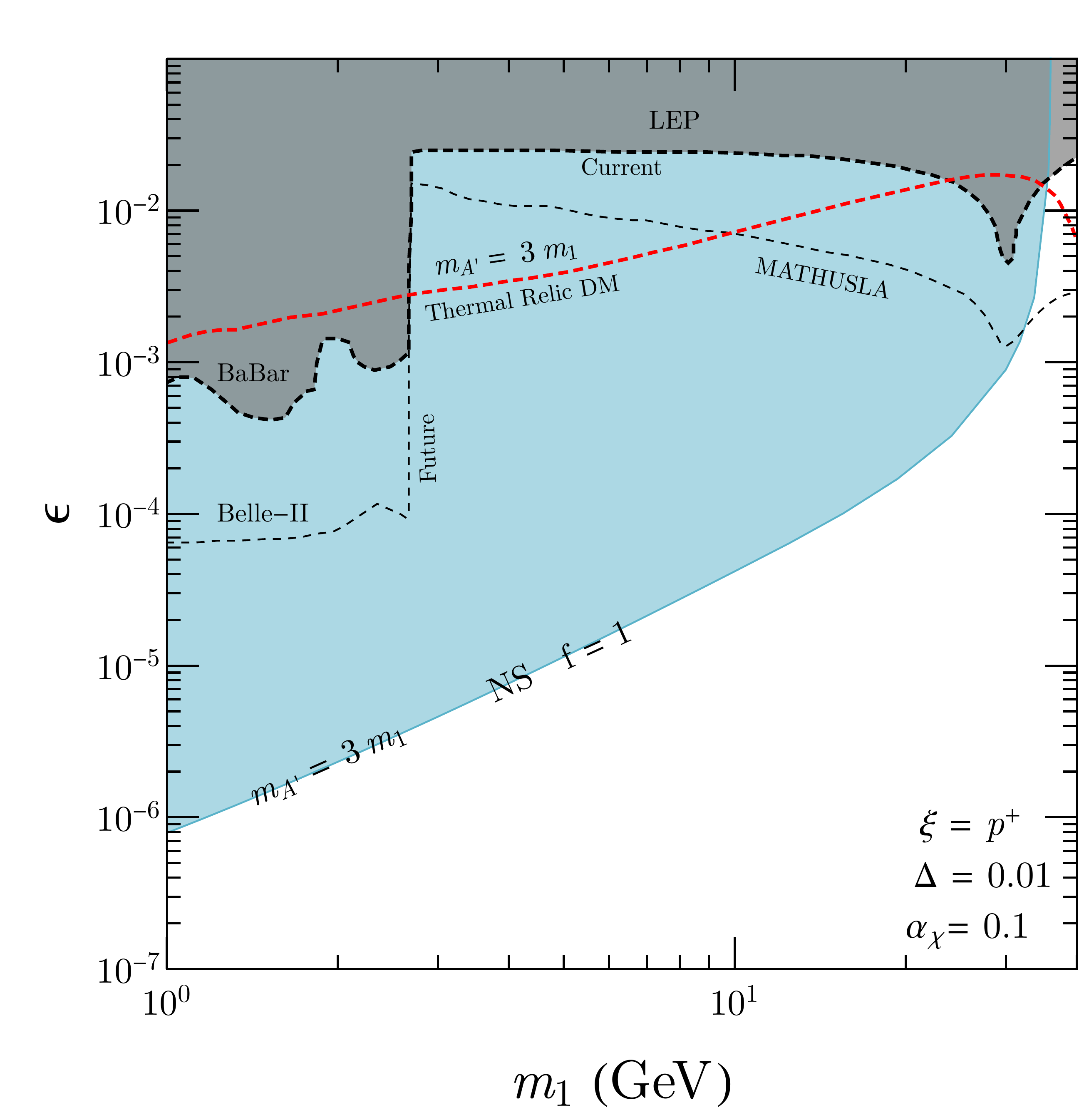}
	\caption{Projected constraints on the kinetic mixing parameter from neutron star heating for inelastic dark matter, assuming portal interactions with the proton target, where $\Delta=0.03$ (cyan-shaded, left panel) and $0.01$ (cyan-shaded, right panel). The rest is the same as Fig.~\ref{fig:heavymed}. The projected bound for FASER~\cite{Feng:2017uoz,Berlin:2018jbm} is also shown as an example of potential of the LHC to probe this parameter space in the near future.}
	\label{fig:heavymed1} 
\end{figure}

Fig.~\ref{fig:heavymed} (left panel) shows the projected constraints on the kinetic mixing parameter $\epsilon$ from neutron stars with dark matter scattering off ultrarelativistic electrons for $m_{A'}=3m_1$ (dark yellow-shaded) and $m_{A'}=(2+10^{-6})m_1$ (dark yellow-shaded $+$ light yellow-shaded extension), where we fix $\Delta=(m_2-m_1)/m_1=0.1$. For the dark matter mass below $m_1=1.6~$GeV, the projected constraints from the heating can be much stronger than the current upper limits from the BaBar experiment~\cite{BaBar:2017tiz} (boundary of the gray-shaded region), as well as those expected from the future LDMX~\cite{Berlin:2018bsc,Berlin:2018pwi,LDMX:2018cma} and Bell-II~\cite{Belle-II:2018jsg} experiments (dashed-light black). In addition, the heating reach can cover the entire parameter regime relevant for the thermal relic dark matter (dashed-red). When $m_{A'}$ approaches $2m_1$, the resonant effect enhances the co-annhilation $\chi_1\chi_2\rightarrow A'^{*}_\mu\rightarrow\zeta\bar{\zeta}$. Accordingly $\epsilon$ needs to be reduced to obtain the correct relic density. For $m_{A'}=(2+10^{-6})m_1$, the required $\epsilon$ value is about two orders of magnitude less than that for $m_{A'}=3m_1$. In this ``fine-tuned" case, the model can avoid the bounds from the future collider experiments, but it is still subject to potential constraints from neutron star heating. In fact, the projected constraints from the heating become stronger for a light dark photon, as expected. 

Fig.~\ref{fig:heavymed} (right panel) shows the projected constraints from neutron stars with scattering off protons for $m_{A'}=3m_1$ (dark cyan-shaded) and $m_{A'}=2m_1$ (dark cyan-shaded $+$ light cyan-shaded extension). For the dark matter mass below $m_1=2.6~$GeV, we again see that the projected constraints from neutron stars can be stronger than the current upper limits by a factor of $\sim10^3$ from the LSND~\cite{LSND:2001akn,deNiverville:2011it}, E137~\cite{Bjorken:1988as,Batell:2014mga}, and BaBar~\cite{BaBar:2017tiz} experiments (boundary of the gray-shaded region). It can be stronger by a factor of $\sim10^2$ compared to the future projected limits from LDMX ~\cite{Berlin:2018bsc,Berlin:2018pwi,LDMX:2018cma}, JSNS$^2$~\cite{Jordan:2018gcd}, SeaQuest~\cite{Berlin:2018pwi}, and Belle-II~\cite{Belle-II:2018jsg} experiments (dashed-light black). We note that the region for the thermal relic with $m_{A'}=3m_1$ has been entirely excluded by the joint constraints from the collider experiments involving both electron and proton beams. For $m_{A'}=(2+10^{-6})m_1$, the majority of the thermal relic region can avoid constraints from the future terrestrial searches, but it can be constrained from neutron stars.

Fig.~\ref{fig:heavymed1} demonstrates the projected constraints for lower relative mass gaps, $\Delta=0.03$ (left panel) and $0.01$ (right panel), where we focus on proton targets. As $\Delta$ is lowered from $0.1$ to $0.03$, the future experimental limits on $\epsilon$ are weakened by a factor of $10$. On the other hand, the constraints from neutron stars expand to include $3$ to $4$ times higher dark matter masses, while maintaining the projected limits on $\epsilon$. For lower $\Delta$, terrestrial searches for long-lived particles or forward searches suffer from the reduction of sensitivity, as the final state is less boosted. In contrast, the constraints from neutron stars become stronger because less energy is needed for upscattering in the star as the mass gap is smaller. We see a similar trend on further lowering $\Delta$ to $0.01$. For electron targets, the heating constraints on $\epsilon$ will be weakened by a factor of $10$ and the $m_1$ reach becomes slightly smaller, compared to those with proton targets shown in Fig.~\ref{fig:heavymed1}, while there are no constraints from MATHUSLA and FASER.

%Therefore, for the brevity we do not explicitly show the plots for the leptophilic case.

 %similar to Fig.~\ref{fig:heavymed} with rest of the description remaining the same

%the effects of varying the relative mass gap . As the $\Delta$ is lowered to $0.03$ (left panel), the future experiment constraints on $\epsilon$ weaken by about an order of magnitude, while the neutron star constraints expand to include $3$ to $4$ times higher DM masses at the same time maintaining the projected limits on $\epsilon$. Terrestrial searches for long lived particles or forward searches suffer the reduction in of sensitivity for lower $\Delta$ values due to the dark particles being not enough long lived or not enough boosted, while the neutron star heating gets stronger due to less energy needed for up-scattering in the star due to lowered mass gap. Lowering the $\Delta$ further to $0.01$ (right panel) leads to furthering of the same effects. In the case of leptophilic interactions, the heating bounds on $\epsilon$ will be weaker by about an order of magnitude compared to proton target heating bounds and $m_1$ reach slightly smalled, similar to Fig.~\ref{fig:heavymed} with rest of the description remaining the same. Therefore, for the brevity we do not explicitly show the plots for the leptophilic case.

We have seen that neutron star constraints can be superior in probing inelastic dark matter, compared to current and future terrestrial collider experiments. They are particularly more powerful for intermediate $\Delta$ values between $10^{-5}$ and $10^{-2}$, where both the accelerator and direct detection searches are not good at accessing the cosmologically interesting parameter space that has not yet been excluded. Our results in Fig.~\ref{fig:heavymed1} are based on the model where the dark photon couples to both leptons and quarks in the Standard Model. They can be easily extended to other models like inelastic leptophilic dark matter models, where neutron stars could be the best target for searching such interactions. As depicted in Fig.~\ref{fig:heavymed}, the terrestrial bounds for the leptophilic case will be considerably reduced, while the projected heating constraints are only slightly weaker compared to the non-leptophilic case. We also note that that the parameter region $1<m_{A'}/m_1<2$ is studied in detail in~\cite{Fitzpatrick:2021cij}. Based on the results presented in this section, we find that the open gaps in the $\epsilon-m_1$ plane pointed out~\cite{Fitzpatrick:2021cij} can be probed by neutron stars.

\section{Constraining inelastic dark matter with a light mediator}
\label{sec:5}

\begin{figure}[tp]
	\centering
	\includegraphics[width=.48\textwidth]{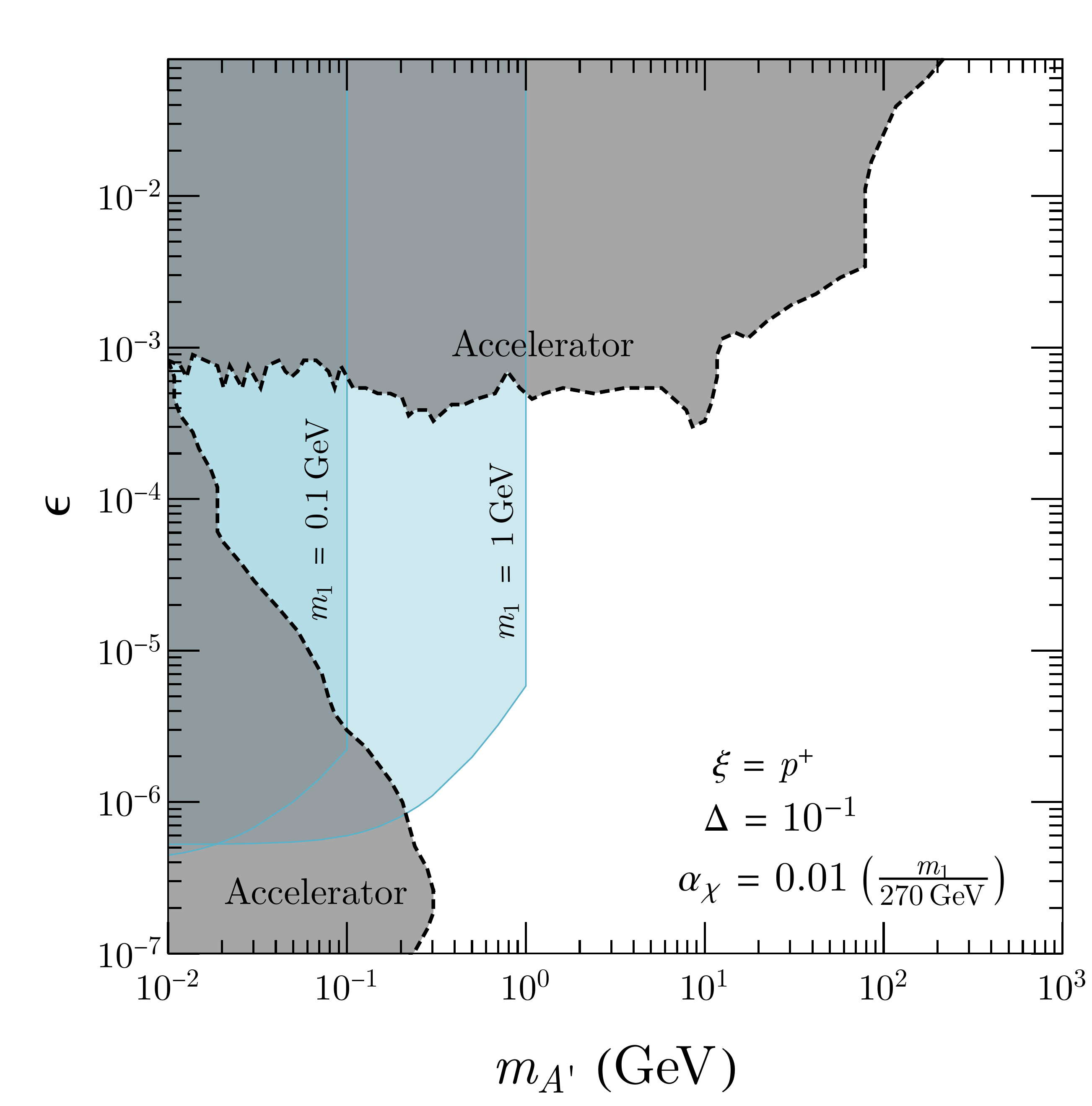}
	\includegraphics[width=.48\textwidth]{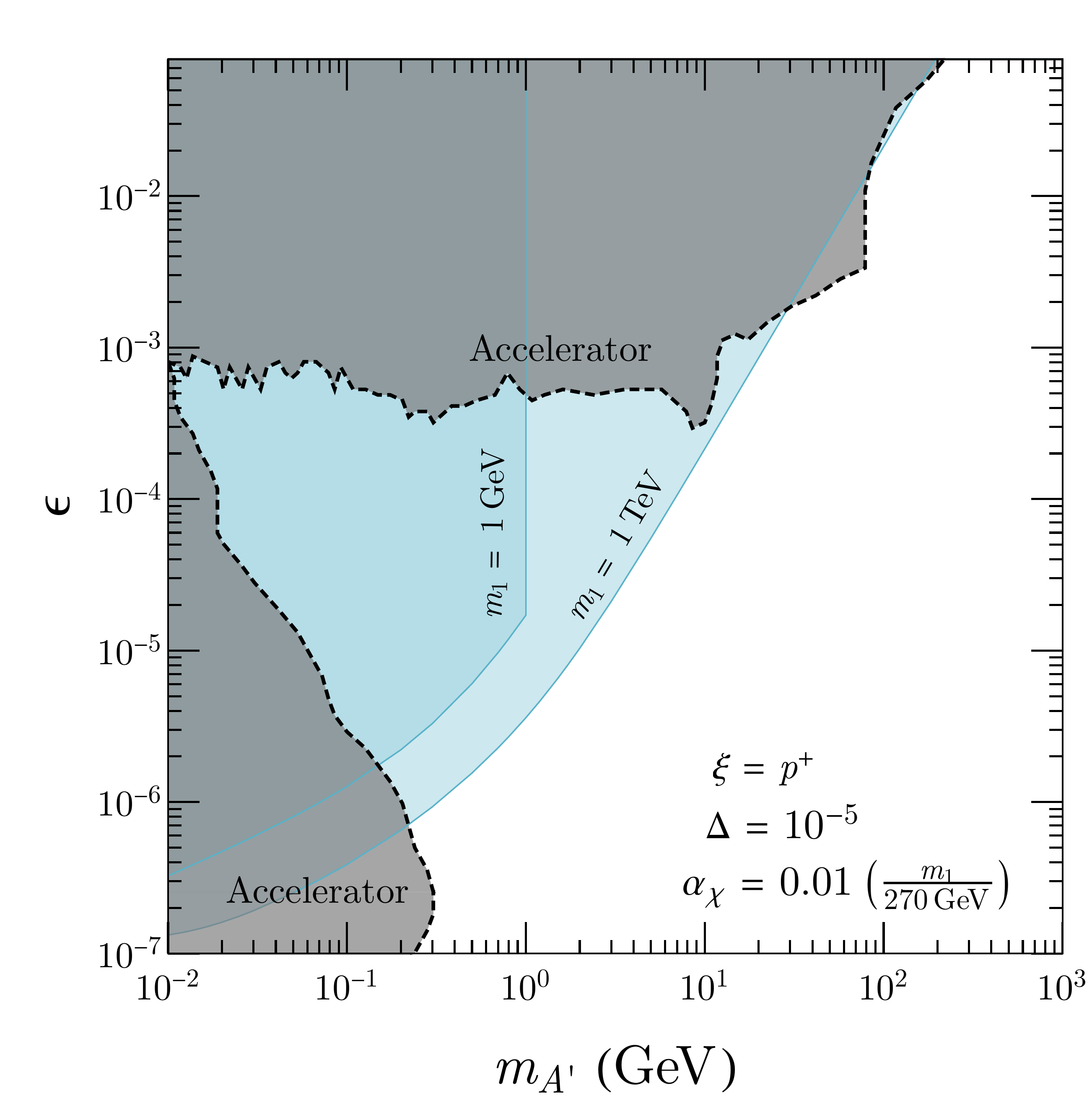}
	\caption{Projected constraints from neutron star heating for inelastic dark matter assuming portal interactions with the proton target for fixed dark matter masses as denoted, assuming $\Delta=10^{-1}$ (cyan-shaded, left panel) and $10^{-5}$ (cyan-shaded, right panel). The dark fine structure constant is fixed to be $\alpha_\chi=0.01\left(m_1/270\,\text{GeV}\right)$, consistent with the relic density constraint. The gray-shaded region is excluded by accelerator experiments~\cite{Andreas:2012mt,BaBar:2014zli,Anastasi:2015qla,NA482:2015wmo,LHCb:2017trq,Lin:2019uvt}. }
	\label{fig:secluded} 
\end{figure}

Now we consider inelastic dark matter with a light mediator. For simplicity, we consider a model with the same Lagrangian as in Eq.~\ref{eq:model}, but assume a small dark photon mass $m_{A'}<m_1$. In this case, the dark matter abundance is set by the s-wave annihilation process $\chi_1\chi_1\rightarrow A'_\mu A'_\mu$~\cite{Boehm:2003hm,Pospelov:2007mp,Feng:2008ya,Arkani-Hamed:2008hhe,Feng:2009mn,Tulin:2013teo}. To avoid stringent constraints from indirect detection searches~\cite{Bringmann:2016din,Cirelli:2016rnw,Baldes:2020hwx}, we further assume that there exists a massless fermion in the dark sector to which the dark photon predominantly decays, see, e.g.,~\cite{Huo:2017vef}. Fig.~\ref{fig:secluded} shows various constraints in the $\epsilon\textup{--}m_{A'}$ plane for the model with a light mediator, where we fix $\alpha_\chi$ to the values required for the correct relic density. The gray regions denote the $\epsilon$ values already excluded by accelerator or astrophysical observations~\cite{Lin:2019uvt}, irrespective of the dark matter model under consideration. The left panel in Fig.~\ref{fig:secluded} shows projected constraints from neutron stars for $m_1=100$~MeV (dark cyan-shaded) and $m_1=1$~GeV (dark cyan-shaded $+$ light cyan-shaded extension) in the case of $\Delta=0.1$. In the right panel, we show the projected constraints in the case of  $\Delta=10^{-5}$ for $m_1=1$~GeV (dark cyan-shaded) and $m_1=1$~TeV (dark cyan-shaded $+$ light cyan-shaded extension). Both $\Delta$ values are unreachable by direct detection experiments. The $m_{A'}$ reach is constrained not by the kinematics, but due to the model assumptions, because the approximate maximum mass gap allowed by $\Delta$ value as indicated in Table~\ref{tab:dmax} is always satisfied for $m_1<m_{A'}$.

For the model of inelastic dark matter with a light mediator, dark matter may have strong self-interactions. Although the upscattering process $\chi_1+\chi_1\rightarrow\chi_2+\chi_2$ can be kinematically forbidden in dark matter halos of dwarf galaxies, the elastic scattering process $\chi_1+\chi_1\rightarrow\chi_1+\chi_1$ can still occur and its corresponding scattering cross section receives nonperturbative quantum corrections~\cite{Zhang:2016dck,Blennow:2016gde,Alvarez:2019nwt}. The self-interactions can thermalize the inner regions of dark matter halos and affect the distribution of dark matter in galaxies. Recent studies have shown that such a scenario can be favored in explaining kinematic measurements of stars and gas particles in dwarf galaxies, see~\cite{Tulin:2017ara,Adhikari:2022sbh}. The advantage of considering inelastic self-interacting dark matter is that it can evade stringent direct detection constraints~\cite{PandaX-II:2018xpz,PandaX-II:2021lap} if the mass splitting between $\chi_1$ and $\chi_2$ is sufficiently large such that $\chi_1+\xi\rightarrow\chi_2+\xi$ is forbidden. %However, we show that the scenario can be strongly constrained with neutron stars. 

\begin{figure}[t]
	\centering
	\includegraphics[width=.48\textwidth]{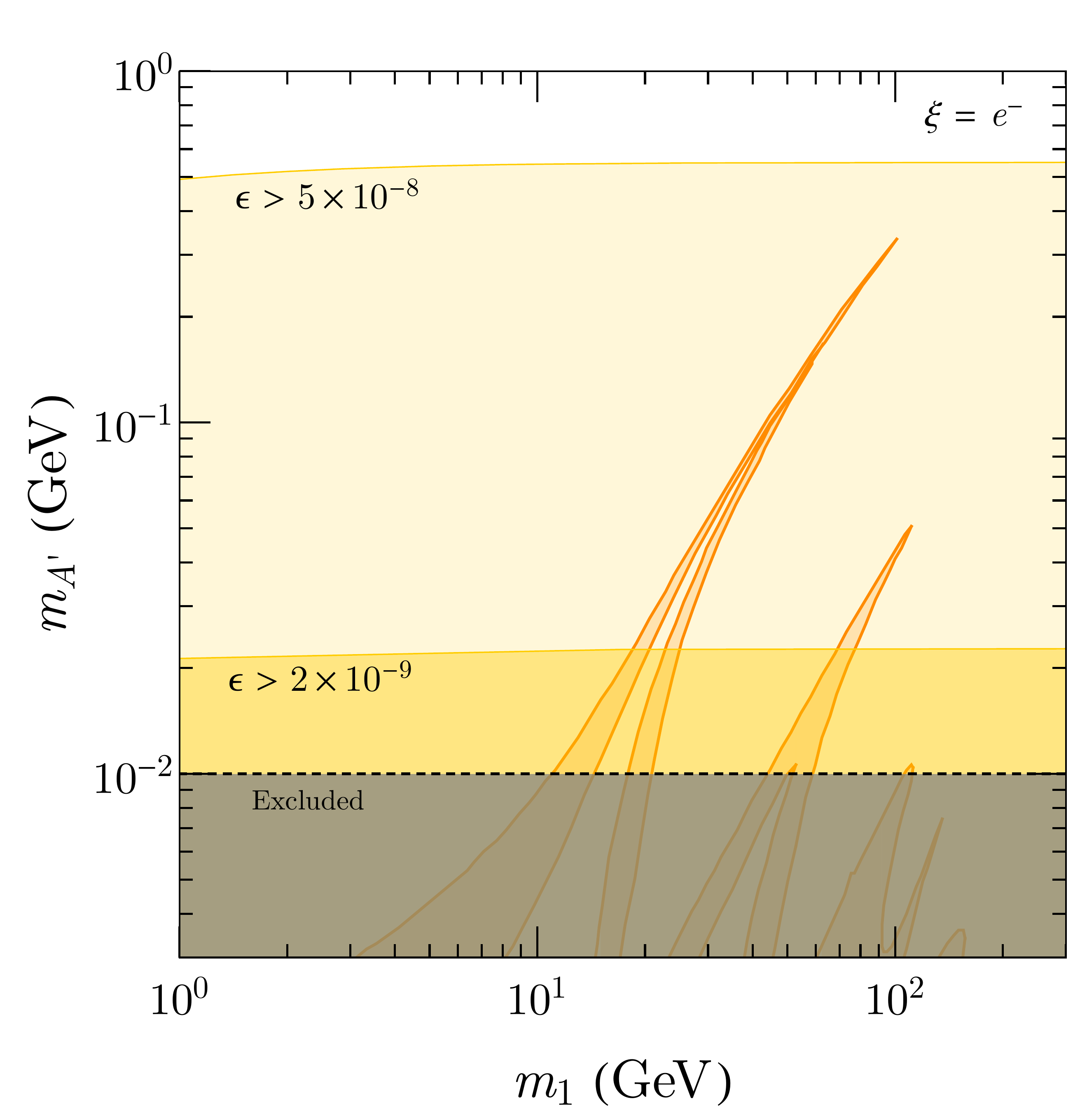}
	\includegraphics[width=.48\textwidth]{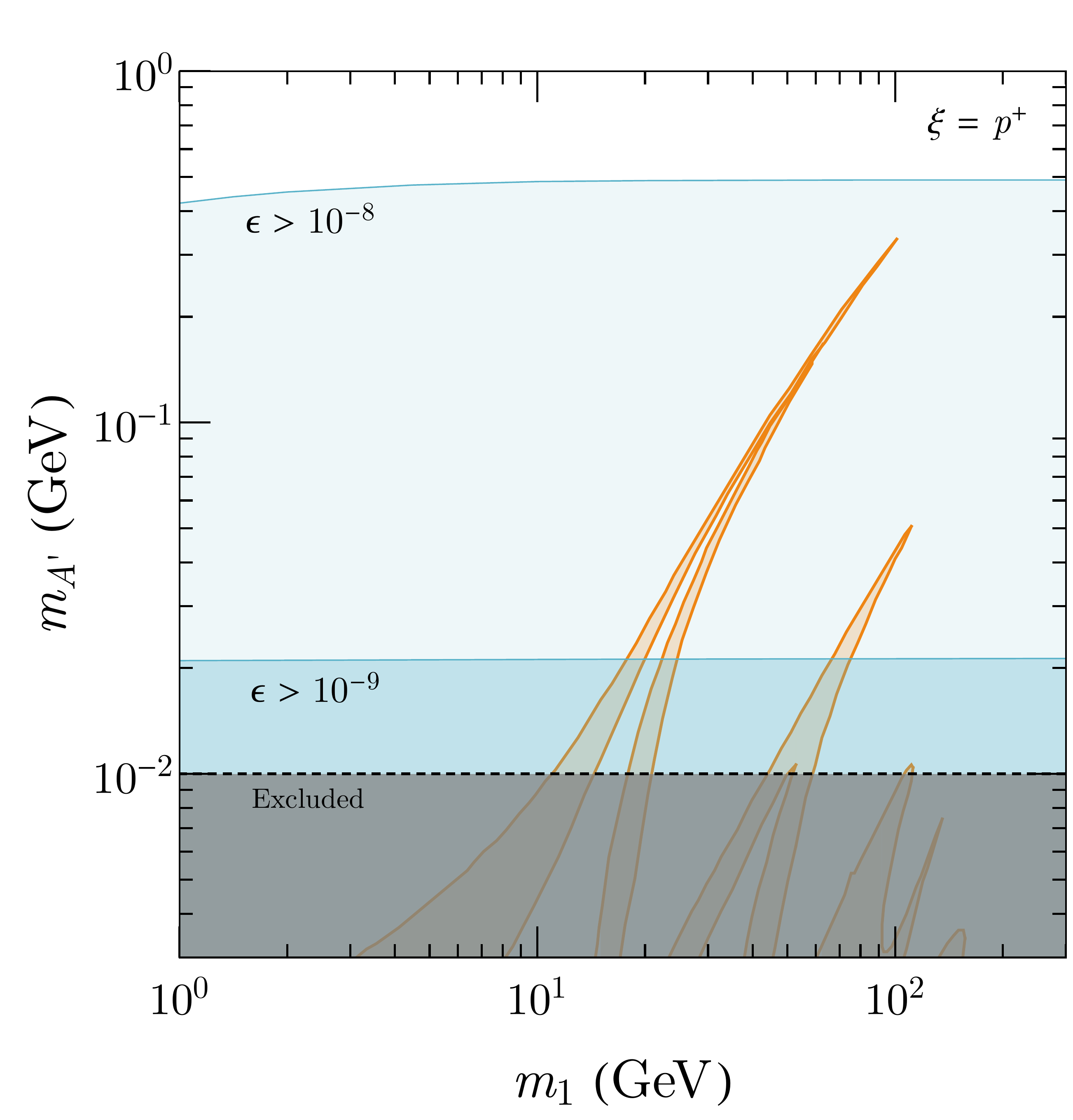}
	\caption{Projected constraints on kinetic mixing parameter $\epsilon$ from neutron star heating in the $m_{A'}\textup{--}m_1$ plane for an inelastic self-interacting dark matter model with electron (yellow-shaded, left panel) and proton (cyan-shaded, right panel) targets. %assuming the kinetic mixing parameter to be $4\times10^{-9}$ and $4\times10^{-8}$ [check]. 
	The mass splitting is set to be $\delta m=1~{\rm MeV}$ that evades the constraints from terrestrial direct detection experiments. The dark fine structure constant is fixed as $\alpha_\chi=0.01$. In the orange region, the self-scattering cross section per mass predicted by the model satisfies the condition $0.5\,{\rm cm^{2}/g} \leq \sigma ({\chi_{1}\chi_{1} \rightarrow \chi_{1} \chi_{1}}) / m_1 \leq 5 \,{\rm cm^{2}/g}$~\cite{Blennow:2016gde}, which is relevant for producing dark matter density cores in dwarf galaxies. Gray region shows excluded parameter space due to accelerator and supernova constraints~\cite{Andreas:2012mt,BaBar:2014zli,Anastasi:2015qla,NA482:2015wmo,LHCb:2017trq,DeRocco:2019njg,Lin:2019uvt}.}
    \label{fig:SIDM} 
\end{figure}

Refs.~\cite{Blennow:2016gde,Alvarez:2019nwt} numerically solve the Schr\"odinger equation to find favored regions in $m_{A'}-m_1$ plane for various values of $\alpha_\chi$ and $\delta m$. Fig.~\ref{fig:SIDM} shows the region where the self-scattering cross section per mass is in the range $0.5~{\rm cm^2/g}\leq\sigma(\chi_1\chi_1\rightarrow\chi_1\chi_1)/m_1\leq5~{\rm cm^2/g}$ for $\alpha_\chi=0.01$ and $\delta m=1~{\rm MeV}$ (orange), taken from~\cite{Blennow:2016gde}. The self-scattering cross section in this range is favored in explaining diverse dark matter distributions of isolated dwarf galaxies and an even larger value would also work, see~\cite{Kamada:2016euw,Ren:2018jpt} for details. Since the mass gap is $\delta m=1~{\rm MeV}$, the model evades bounds from conventional direct detection searches. However, %the parameter space can be constrained from the consideration of neutron star heating as 
kinetic energy of dark matter particles approaching a neutron star is high enough to overcome the mass gap, thus the model is subject to projected constraints from neutron stars.% $\delta m=1~{\rm MeV}$.

Fig.~\ref{fig:SIDM} shows the $m_{A'}-m_1$ plane of the parameter space that can be probed by neutron stars. The values indicated on the boundary of each yellow- and cyan-shaded region in Fig.~\ref{fig:SIDM} correspond to the minimum $\epsilon$ value that can be probed in the respective region. The left panel represents the projected constraints due to electron targets, while the right panel corresponds to proton targets. The entire region corresponding to $0.5~{\rm cm^2/g}\leq\sigma(\chi_1\chi_1\rightarrow\chi_1\chi_1)/m_1\leq5~{\rm cm^2/g}$ %for $m_{A'}\lesssim20~{\rm MeV}$ 
could be probed with neutron star heating for $\epsilon>10^{-8}$ when dark matter couples to protons and $\epsilon>5\times 10^{-8}$ for dark matter that couples to only electrons. The projected constraints are largely insensitive to $m_1$. This behavior is due to the interaction cross section being very close to saturation limit for the values of $m_{A'}$ and $m_1$ under consideration in Fig.~\ref{fig:SIDM}.

In Fig.~\ref{fig:SIDM}, the dark fine structure constant is fixed to $\alpha_\chi=0.01$. Therefore, the orange region corresponds to $\chi_1$ forming large percentage of thermal relic density only in the region $\mathcal{O}(100)$~GeV. A detailed analysis can be performed similar to work in~\cite{Alvarez:2019nwt} to map thermal relic dark matter that satisfies all the small-scale structure constraints on the parameter space in Fig.~\ref{fig:secluded}. For the purpose of illustration in Fig.~\ref{fig:SIDM}, we remain agnostic to the cosmological origin of the dark matter and show the potential of neutron star heating to probe self-interacting dark matter compatible parameter space for fixed $\alpha_\chi=0.01$ as in~\cite{Blennow:2016gde}.

We have shown that neutron stars can provide an excellent probe of the parameter space that is not covered by the terrestrial and other astrophysical searches for the models with a light mediator. Inelastic dark matter models with dark photon mediators can be probed for $\epsilon\ge \mathcal{O}(10^{-7})$ for a large range of $m_1$ from $10$~MeV to $10$~TeV, when $\alpha_\chi$ is fixed to a value that makes the dark matter undergo a dark freeze-out with correct thermal relic density.%The same models can also provide an excellent way of explaining some of the small scale structure problems while evading direct and indirect detection. 
We have also demonstrated the potential of neutron star heating to probe $\epsilon\ge \mathcal{O}(10^{-9})$ in the parameter space relevant for strong dark matter self-interactions.

\section{Discussion and conclusions}
\label{sec:6}

The dense environment of neutron stars makes these compact objects an excellent target for probing dark matter interactions with the Standard Model. In this work, we have explored the capture of inelastic dark matter in neutron stars and the associated heating effect. We studied kinematics of the inelastic scattering process between quasirelativistic dark matter particles and ultrarelativistic targets. We also derived analytical expressions for the maximal mass gap allowed for the scattering to occur. These results were then implemented in a fully relativistic formalism for calculating the capture rate of inelastic dark matter in a neutron star. We applied the formalism to various models and obtained potential constraints.

We considered simplified models where inelastic dark matter couples to the Standard Model via an effective contact operator and a vector portal interaction. For much of the parameter space, the projected constraints from neutron stars complement and surpass direct detection experiments searching for inelastic dark matter. Furthermore, we considered a concrete model, where psuedo-Dirac dark matter couples to the Standard Model via kinetic mixing, and found that the heating constraints on the mixing parameter can be $4\textup{--}5$ orders of magnitude stronger than those from collider and beam dump experiments. We find that the neutron star constraint could be most powerful for an intermediate mass gap $\Delta$ values between $10^{-5}$ and $10^{-2}$, compared to terrestrial experiments. In addition, we showed that neutron stars can probe parameter space of inelastic self-interacting dark matter, where terrestrial direct detection bounds are evaded.  
%Supernova cooling and accelerator bounds exclude a portion of the parameter space, but there remains parameter space which can only be probed using NS heating. 

We have focused on a vector-vector contact operator and a vector mediator, but it would be straightforward to extend the analysis to all possible dimension 5 and 6 operators for bosonic and fermionic dark matter, respectively, as in the elastic case~\cite{Joglekar:2020liw}. The extension to the inelastic case for those operators will follow a similar curve to the elastic case in the $\Lambda-m_1$ plane until $\delta m$ exceeds the allowed maximum mass gap. There will be a sharp drop-off in the reach at $m_1$ where the condition is violated, thus losing the sensitivity for higher $m_1$ when $\Delta$ is fixed. The mass gaps shown in Table~\ref{tab:dmax} are independent of the type of operator or the mediator. Hence the dark matter masses at which the inelastic reach sharply deviates from the elastic ones will be similar to those in the vector portal or vector-vector contact operators presented here for given $\Delta$ or $\delta m$. 

In deriving the projected constraints, we have adopted a simplified approach to model the neutron star. There are several uncertainties associated with this approach as discussed in~\cite{Joglekar:2019vzy,Joglekar:2020liw}. They mainly stem from uncertainties in the density profile of the contents of the star, the equation of state and the velocity distribution of dark matter particles in the halo, as well as departure from the free Fermi gas approximation. Ref.~\cite{Joglekar:2019vzy,Joglekar:2020liw} estimated that in the elastic case these uncertainties could lead to a very small $\mathcal{O}(1)$ factor change in the projected limits on the cutoff scale $\Lambda$ for both relativistic and nonrelativistic targets, partly because $\Lambda$ scales as $\Lambda\propto f^{1/4}$. Accordingly, for the projected limits on the coupling constants $g_\chi g_\xi$ and $\epsilon$, the uncertainties can translate into a small to medium $\mathcal{O}(1)$ factor, since they scale as $f^{1/2}$. We expect these estimates to be applicable to the inelastic case in this work as well. 

We have shown that the neutron stars can offer an independent, complementary way to search for inelastic dark matter for a wide range of dark matter masses, including the parameter space that is inaccessible in terrestrial experiments. In the future, we could further reduce the theoretical uncertainties with better modeling. %Observations of $\mathcal{O}(10^{9})$ year old neutron stars at temperature of $1500\textup{--}2500$~K could be enough to discover particle dark matter or else put strong constraints on the model. 
Observations of $\mathcal{O}(10^{9})$ year old neutron stars at temperatures of $1500$–$2500$~K would be essential for carrying out the proposed search. This is exciting as the next generation radio telescopes, such as FAST~\cite{Nan:2011um}, CHIME~\cite{Ng:2017djg}, and SKA~\cite{Carilli:2004nx}, and (upcoming) infrared telescopes like JWST~\cite{Gardner:2006ky}, TMT~\cite{Skidmore:2015pvw}, and EELT~\cite{Andersen:2003} can potentially discover such a neutron star.

\section*{Acknowledgments}
%AJ's work at LAPTh was supported by the Labex grant ENIGMASS. HBY was supported by the U.S. Department of Energy under Grant No. de-sc0008541.
This work was supported in part by the Labex grant ENIGMASS (AJ) and the
U.S. Department of Energy under Grant No. de-sc0008541 (GA, MPM, HBY).

\appendix

\section{Details of Kinematics}
\label{sec:AppA}

%Let $\delta m$ be the mass gap between 2 states of the inelastic DM. {\color{orange}

We derive analytical expressions for the maximum mass gap $\delta m_\text{max}$ below which the scattering can take place for given $\beta_\text{esc}$, up to first order in the small parameters defined as ratios of various mass and momentum scales in the problem. With obtained $\delta m_\text{max}$, we can estimate the $m_1$ reach of the projected neutron star constraints. These formulas are approximate and the predicted $m_1$ reach agrees with that from the numerical calculation within $\mathcal{O}(10\%)$. %But the formulas capture the scalings with various mass and momentum parameters in the problem.

\subsection {Ultrarelativistic targets and light dark matter}

%Here we consider the case with ultrarelativistic targets $\left(p_\text{\tiny{F}}\gg m_\xi\right)$ and light dark matter $\left(p_\text{\tiny{F}}> m_1\right)$. There are three sub-cases: (i) Light-ish dark matter, (ii) Medium-light dark matter $(m_\xi>m_1>m^2_\xi/p_F)$, (iii) Very-light dark matter $m_\xi^2/p_F}>m_1$ as outlined in~\cite{Joglekar:2020liw}. We introduce the following three small parameters: $x=m_1/p_\text{\tiny{F}}$, $y=\delta m/p_\text{\tiny{F}}$, and $z=m_\text{\tiny{T}}/p_\text{\tiny{F}}$. In the following subsections, we make use of the second condition in Eq.~\ref{eqn:conds} which happens to be stronger than the first condition, to estimate maximum possible $\delta m$ for given $\beta_\text{esc}$.

%$(m_\xi^2/p_F}>m_1)$

We consider the ultrarelativistic $(p_\text{\tiny{F}}\gg m_\xi)$ and light dark matter $(p_\text{\tiny{F}}> m_1)$. There are three cases: (i) Light-ish dark matter $(p_\text{\tiny{F}}>m_1>m_\xi)$, (ii) Medium-light dark matter $(m_\xi>m_1>m_\xi^2/p_\text{\tiny{F}})$, and (iii) Very-light dark matter $(m_\xi^2/p_\text{\tiny{F}}>m_1)$ as outlined in~\cite{Joglekar:2020liw}. We introduce the following three small parameters : $x=m_1/p_\text{\tiny{F}}$, $y=\delta m/p_\text{\tiny{F}}$, and $z=m_\xi/p_\text{\tiny{F}}$. In the following subsections, we make use of the second condition in Eq.~\ref{eqn:conds}, which is stronger than the first condition for light dark matter, to estimate maximum possible $\delta m$ for given $\beta_\text{esc}$.

\subsubsection{Light-ish dark matter}
%Since $z\ll x\ll 1$, we first expand in $y$, $z$ and $x$ sequentially, Retaining terms to apparent first orders, we get
%\begin{align}
%\Delta E_\text{\tiny{NS}}+E_\xi-E_\xi^\text{F}&\approx p-p_\text{\tiny{F}}-\frac{\left(1-\cos\psi\right)}{2}p+\left(-\cos\alpha\right)\beta_\text{esc}\left(\sin\theta\right)\sin\psi\sqrt{\frac{\gamma_\text{esc}p_\text{\tiny{F}}p}{2\left(1-\beta_\text{esc}\cos\theta\right)}}\sqrt{x}\notag\\
%&\quad +\frac{1+\beta_\text{esc}^2-2\beta_\text{esc}\cos\theta}{4\left(1-\beta_\text{esc}\cos\theta\right)}\gamma_\text{esc}\left(1-\cos\psi\right)p_\text{\tiny{F}}x\notag\\
%&\quad +\left[\frac{-(1+\cos\psi)p_\text{\tiny{F}}}{2\gamma_\text{esc}\left(1-\beta_\text{esc}\cos\theta\right)}+\left(\cos\alpha\right)\beta_\text{esc}\left(\sin\theta\right)\sin\psi\sqrt{\frac{p_\text{\tiny{F}}^3}{2\gamma_\text{esc}p\left(1-\beta_\text{esc}\cos\theta\right)^3}}\sqrt{x}\right.\notag\\
%&\left.\quad-\frac{1+\beta_\text{esc}^2-2\beta_\text{esc}\cos\theta}{4\left(1-\beta_\text{esc}\cos\theta\right)^2}\,\frac{p_\text{\tiny{F}}^2}{p}\left(1-\cos\psi\right)x\right]y\label{eqn:lightish:full}
%\end{align}
%Note that expressions like $(1-\cos\psi)$ or $\sin\psi$ are not without cost. 
From Eq.~D.20 in~\cite{Joglekar:2020liw}, we know that 
\begin{align}
1-\cos\psi<\frac{2\beta_\text{esc}^2\sin^2\theta\cos^2\alpha}{(1-\beta_\text{esc}\cos\theta)^2}\left[z^2+2x\left(\gamma_\text{esc}(1-\beta_\text{esc}\cos\theta)+z^2\frac{p_\text{\tiny{F}}^2}{p^2}\mathcal{O}(x^2)\right)+\mathcal{O}(x^2)\right].\label{eqn:angle:max:origin}
\end{align}
Since $z\ll x$, this reduces to
\begin{align}
1-\cos\psi<\frac{4\beta_\text{esc}^2\gamma_\text{esc}\sin^2\theta\cos^2\alpha}{1-\beta_\text{esc}\cos\theta}\,x.\label{eqn:reducedcos}
\end{align}
%Therefore, for the light-ish dark matter case, up to apparent combined first order in $x,y,z$, we get
%\begin{align}
%1-\cos\psi<\frac{4\beta_\text{esc}^2\gamma_\text{esc}\sin^2\theta\cos^2\alpha}{(1-\beta_\text{esc}\cos\theta)}x;\quad \sin\psi<2\sqrt{2}\,\beta_\text{esc}\sin\theta\cos\alpha\sqrt{\frac{\gamma_\text{esc}}{(1-\beta_\text{esc}\cos\theta)}}\sqrt{x}. \label{eqn:angle:max}
%\end{align}
Therefore, we use ansatz that $1-\cos\psi=w x$, where $w$ is a constant for given $\cos\alpha$ and $\sin\theta$. This is justified since we are interested in deriving an expression for maximum $\delta m$ only up to combined first order in $x$, $y$, and $z$. Based on Eq.~\ref{eqn:reducedcos}, we expect
\begin{align}
0<w<4\beta^2_\text{esc}\gamma_\text{esc}\sin^2\theta\cos^2\alpha/(1-\beta_\text{esc}\cos\theta).\label{eqn:expect}
\end{align}
It is easiest to fulfill the second condition of Eq.~\ref{eqn:conds} for $p$ very close to the Fermi surface. Using the ansatz, we evaluate $\Delta E_\text{\tiny{NS}}+E_\xi-E_\xi^\text{F}$ in the limit $p\rightarrow p_\text{\tiny F}$. As $z\ll x$, we expand in $y$, $z$ and $x$ sequentially and only retain up to the combined first-order terms to get 
\begin{align}
\Delta E_\text{\tiny{NS}}+E_\xi-E_\xi^\text{F}&\approx\frac{p_\text{\tiny F}}{2}\left(-w-2\sqrt{w}\,\beta_\text{esc}\cos\alpha \sin\theta\sqrt{\frac{\gamma_\text{esc}}{1-\beta_\text{esc}\cos\theta}}\right)x-\frac{p_\text{\tiny F}}{\gamma_\text{esc}\left(1-\beta_\text{esc}\cos\theta\right)}y\label{eqn:lightish:energy:trans}
\end{align}
The term on the right-hand side is maximized for $\cos\alpha=-1$ and $\sin\theta=1$. Using these we find that the expression is maximized for 
\begin{align}
w=\beta^2_\text{esc}\gamma_\text{esc}
\end{align}
This value is consistent with the expectation in Eq.~\ref{eqn:expect}. Substituting it in Eq.~\ref{eqn:lightish:energy:trans}, the second condition in Eq.~\ref{eqn:conds} becomes
\begin{align}
\delta m<\frac{1}{2}\beta^2_\text{esc}\gamma_\text{esc}^2 m_1.
\end{align}

The upper bound on the $y$ and the corresponding $\psi$ in the medium-light dark matter case is same as the light-ish dark matter case, so we don't elaborate that further.

\subsubsection{Very light dark matter}
In this case, we have $x\ll z^2\ll z \ll 1$. Therefore, on retaining the largest term, Eq.~\ref{eqn:angle:max:origin} reduces to
\begin{align}
1-\cos\psi<\frac{2\beta_\text{esc}^2\sin^2\theta\cos^2\alpha}{(1-\beta_\text{esc}\cos\theta)^2}z^2.
\end{align}
We can use ansatz $1-\cos\psi=w z^2$, and the maximum possible range of $w$ is 
\begin{align}
0<w<2\beta^2_\text{esc}\sin^2\theta\cos^2\alpha/(1-\beta_\text{esc}\cos\theta)^2.\label{eqn:expect:verylight}
\end{align}
Similar to the previous case, using the ansatz, we evaluate $\Delta E_\text{\tiny{NS}}+E_\xi-E_\xi^\text{F}$ in the limit $p\rightarrow p_\text{\tiny F}$. But now $x\ll z^2\ll z$, so we expand in $y$, $x$ and $z$ sequentially and only retain terms up to the first order in $x$ to get 
\begin{align}
\Delta E_\text{\tiny{NS}}+E_\xi-E_\xi^\text{F}&\approx\gamma_\text{esc}\,p_\text{\tiny F}\left(-w\left(1-\beta_\text{esc}\cos\theta\right)-\sqrt{2\,w}\,\beta_\text{esc}\cos\alpha \sin\theta\right)x-\frac{p_\text{\tiny F}}{\gamma_\text{esc}\left(1-\beta_\text{esc}\cos\theta\right)}y\label{eqn:verylight:energy:trans}
\end{align}
The term on the right-hand side is maximized for $\cos\alpha=-1$ and $\sin\theta=1$. Using these we find that the expression is maximized for 
\begin{align}
w=\beta^2_\text{esc}/2
\end{align}
This value is consistent with the expectation in Eq.~\ref{eqn:expect:verylight}. Substituting it in Eq.~\ref{eqn:verylight:energy:trans}, the second condition in Eq.~\ref{eqn:conds} becomes
\begin{align}
\delta m<\frac{1}{2}\beta^2_\text{esc}\gamma_\text{esc}^2 m_1.
\end{align}

\subsection {Ultrarelativistic targets and heavy dark matter}
In this case, we have $p_\text{\tiny{F}}\gg m_\xi$ and $p_\text{\tiny{F}}< m_1$. The second condition in Eq.~\ref{eqn:conds} is still the limiting one. So we again evaluate $\Delta E_\text{\tiny{NS}}+E_\xi-E_\xi^\text{F}$ at $p\sim p_\text{\tiny{F}}$, because the momentum transfer is the highest when the target has high momentum. We expand in $\delta m/m_1$, $m_\xi/m_1$ and $p_\text{\tiny{F}}/m_1$ sequentially, followed by retaining only the first order terms, to get
\begin{align}
\Delta E_\text{\tiny{NS}}+E_\xi-E_\xi^\text{F}&\approx \beta_\text{esc}\gamma_\text{esc}^2p_\text{\tiny{F}}\left[\left(\beta_\text{esc}-\cos\theta\right)\left(1-\cos\psi\right)-\left(\beta_\text{esc}\cos\alpha\,\sin\theta\sin\psi\right)/\gamma_\text{esc}\right]\notag\\&-\gamma_\text{esc}\left(1+\frac{\beta_\text{esc}\left(1-\beta_\text{esc}\cos\theta\right)\left[(-\cos\alpha)\sin\theta\sin\psi/\gamma_\text{esc}-\left(\beta_\text{esc}-\cos\theta\right)\cos\psi\right]}{\left(1-\beta_\text{esc}\cos\theta\right)^2}\right)\delta m
\end{align} 
This is maximized for head-on collisions as expected. Thus, substituting the corresponding angles $\cos\theta=\cos\psi=-1$, the second condition in Eq.~\ref{eqn:conds} becomes
\begin{align}
\delta m\le 2\beta_\text{esc}\gamma_\text{esc}\,p_\text{\tiny{F}}.
\end{align}

\subsection {Nonrelativistic targets, heavy and light dark matter}
For heavy dark matter, we have $p_\text{\tiny{F}}\ll m_\xi<m_1$. In the extreme nonrelativistic case, we can take  $p_\text{\tiny{F}}\rightarrow 0$. In this limit, the first condition in Eq.~\ref{eqn:conds} is a stronger, because there is no need to knock targets out of their Fermi surface. To evaluate the condition $k'^2_\text{cm}>0$, we solve the quartic in $\delta m$ and find
\begin{align}
\delta m&< -m_1+\sqrt{m_1^2+2\left(m_1^2+\gamma_\text{esc}m_\xi m _1-m_\xi\sqrt{m_\xi^2+2\gamma_\text{esc}m_\xi m_1+m_1^2}\right)}.
\end{align} 
We expand the right-hand side in $m_\xi/m_1$, retain the lowest order, and obtain
\begin{align}
\delta m&<\left(\gamma_\text{esc}-1\right)m_\xi\label{eqn:nrhd}
\end{align}
In the light dark matter case, we expand in $m_1/m_\xi$ and retain the lowest order to get
\begin{align}
\delta m&<\left(\gamma_\text{esc}-1\right)m_1\label{eqn:nrld}
\end{align}
The other inequality in the first condition of Eq.~\ref{eqn:conds}, $k'^2_\text{cm}<k_\text{cm}^2$, gives weaker constraints of $2\gamma_\text{esc}m_\xi$ and $\left(\gamma_\text{esc}-1\right)m_\chi+2m_\xi$ for heavy and light dark matter, respectively. Therefore, Eq.~\ref{eqn:nrhd} and Eq.~\ref{eqn:nrld} set the upper bound on $\delta m$ that is compatible with the scattering.

\section{Recoil Rates and Terrestrial Direct Detection}
\label{sec:AppB}

We briefly discuss the derivation of the direct detection constraints shown in Fig.~\ref{fig:fixedgap} (dashed-blue curves, bottom panels). 
In ground-based detectors, dark matter particles can scatter off the nucleus, and the differential scattering rate is
\begin{equation}
\frac{dR}{dE_R}=n_T\frac{\rho}{m_1}\int\limits_{v_{\rm min}}^{v_{\rm max}}v f(\vec{v},\vec{v_e})\frac{d\sigma}{dE_R}d^3v.
\label{eqn:ER}
\end{equation}
where $E_R$ is recoil energy, $n_T$ the number of target nuclei in a unit mass of the detector, $\rho\approx 0.3~ {\rm GeV/cm^3}$ the local dark matter mass density, and $f(\vec{v},\vec{v_e})$ is the velocity distribution function of dark matter particles, and $d\sigma/dE_R$ the differential cross section for dark matter-nucleus scattering. $\vec{v}$ is the dark matter velocity to the detector and $\vec{v_e}$ is the Earth velocity relative to the distribution (in the Galactic frame). $v_{\rm min}$ and $v_{\rm max}$ are minimum and maximum velocities in the Earth frame, respectively. 

We assume a truncated Maxwell-Boltzmann distribution
\begin{equation}
f\left(v\right)=N\exp\left[\frac{-(v^2+v^2_e+2 vv_e\cos\theta)}{v^2_0}\right],
\end{equation}
where $N$ is a normalization factor, $\theta$ is the angle between $\vec{v}$ and $\vec{v_e}$, and $v_0$ is a characteristic velocity. Assuming the escape velocity of dark matter particles in the Galactic frame to be $v_{\rm esc}$, we can write $N$ as
\begin{equation}
    N=1/\left[\pi^{\frac{3}{2}}v_0^3{\rm erf}\left(v_{\rm esc}/v_0\right)-2v^2_0v_{\rm esc}\exp\left(-v_{\rm esc}^2/v_0^2\right)\right]. 
\end{equation}

The maximum velocity in the Earth frame $v_{\rm max}$ is related to $v_{\rm esc}$ in the Galactic frame as $v^2_{\rm esc}=v^2_{\rm max}+v^2_e+2 v_{\rm max}v_e\cos\theta$. In our analysis, we take $v_{\rm esc}=533~ {\rm km/s}$, $v_0=220~{\rm km/s}$, and $v_e=232~{\rm km/s}$, where we have neglected the relative motion between Earth and Sun as the effect is minor for the purpose of our analysis.

The minimum velocity of dark matter particles for inelastic scattering in the detector is 
\begin{align}
v_{\rm min}\approx \frac{\left(E_{R} m_{N}/\mu+\delta m\right)}{\sqrt{2E_{ R}m_{N}}}, 
\end{align} 
where $m_N$ is the nucleus mass, and $\mu$ is the reduced mass of $m_1$ and $m_{N}$. Depending on the relative magnitude of $v_{\rm min}$ and $v_{\rm esc}-v_e$, the velocity integral in Eq.~\ref{eqn:ER} can be evaluated as%~\cite{Lewin:1995rx}
\begin{equation}
\int\limits_{v_{\rm min}}^{v_{\rm max}} d^3v=2 \pi 
	  \left\{ \begin{array}{ll} 
	  \int\limits_{v_{\rm min}}^{v_{\rm esc}-v_{e}}v^2dv \int\limits^1_{-1}d \cos\theta+\int\limits_{v_{\rm esc}-v_{ e}}^{v_{\rm esc}+v_{ e}}v^2dv \int\limits^{c_*}_{-1}d \cos\theta ~~~~~~~v_{\rm min}<v_{\rm esc}-v_e \\
	  &\\
	\int\limits_{v_{\rm min}}^{v_{\rm esc}+v_{e}}v^2dv \int\limits^{c*}_{-1}d \cos\theta~~~~~~~%\hspace{110}
	  v_{\rm esc}-v_{e}<v_{\rm min}<v_{\rm esc}+v_{e}
	  
	  \end{array}\right.
	  \label{}
\end{equation}
where $c_*=\left(v_{\rm esc}^2-v^2-v^2_{ e}\right)/\left(2vv_{e}\right)$ is the minimum angle for the scattering to occur when $v_{\rm esc}-v_{e}<v_{\rm min}<v_{\rm esc}+v_{e}$, see, e.g.,~\cite{Bramante:2016rdh}. 

We focus on dark matter-proton scattering and recast the results from the XENON1T experiment~\cite{XENON:2018voc}. For the model we consider, the differential scattering cross section is  
\begin{equation}
    \frac{d\sigma}{dE_R} 
     \approx \frac{m_N}{2 \pi v^2}\frac{  g_\chi^2 g_\xi^2 Z^2}{\left(m^2_{\rm \phi}-\delta m^2+ 2m_N E_R\right)^2}F^2\left(E_R\right),
\end{equation}
where $Z$ is the atomic number of the nucleus, and $F^2\left(E_R\right)$ is the nuclear form factor.  The form factor of xenon is~\cite{Vietze:2014vsa}
\begin{equation}
F^2\left(E_R\right)=\frac{e^{-u}}{A^2}\left(A+\sum_{n=1}^5c_n u^n\right)^2
\end{equation}
where $u=q^2b^2/2$, $b^2=m_n^{-1}(45A^{-1/2}-25A^{-2/3})^{-1}~{\rm MeV^{-1}}$, $q^2=2m_N E_R $, $m_n$ is the mass of neutron, and $A$ is the mass number of the atom. The coefficients $c_n$ for the isotope $^{132}_{~54}$Xe are
\begin{equation}
c_1= - 132.841,~c_2= 38.4859,~c_3= - 4.08455,~ c_4=  0.153298,~ c_5= - 0.0013897.
\end{equation}

%For a given set of mode parameters, we integrate Eq.~\ref{eqn:ER} over the range of recoil energies and find the predicted total number of events. Conversely,

For the XENON1T experiment~\cite{XENON:2018voc}, the energy range is $E_R=4.9\textup{--}40.9~{\rm keV}$ for nuclear recoils and the exposure is $278.8~{\rm days}\times1.3~{\rm ton}$. The total number of observed events is consistent with the expected background. %For no observed events 
Thus, we expect 2.3 events as the the upper bound at $90\%$ confidence level, assuming a Poisson distribution. %interval is 2.3 events, since the observed count is poisson distributed. 
Therefore, to derive the corresponding upper bounds on $g_\chi g_\xi$ for the model, we integrate Eq.~\ref{eqn:ER} over the $E_R$ range from XENON1T, multiply the event rate with the exposure, and demand that the resultant number of events to be less than 2.3 events.%total predicted number of events less than $2.3$, corresponding to $90\%$ confidence level~\cite{Bramante:2016rdh}. %We integrate Eq.~\ref{eqn:ER} over the range of recoil energies as in the XENON1T experiment and t , and demand the predicted number of events less than $2.3$.approximately 

%{\color{red}Finally, upper bound on the parameters of a theory, such as coupling strengths can be placed using the total number of observed events. Specifically, for the XENON1T experiment~\cite{XENON:2018voc}, the energy thresholds are $4.9\textup{--}40.9~{\rm keV}$ for nuclear recoils. For an $1.3\times 278.8~{\rm ton.day}$ exposure of liquid xenon, they did not observe significant excess of events over the expected background. At $90\%$ confidence level, this corresponds to $2.8$ events~\cite{Bramante:2016rdh}. }

%In order to find the total number of the events, we need to integrate Eq.~\ref{eqn:ER} over the range of recoil energy of the experiment. {\color{red}Finally, upper bound on the parameters of a theory, such as coupling strengths can be placed using the total number of observed events. Specifically, for the XENON1T experiment~\cite{XENON:2018voc}, the energy thresholds are $4.9\textup{--}40.9~{\rm keV}$ for nuclear recoils. For $1.3\times 278.8~{\rm ton.day}$ exposure of liquid xenon, they did not observe significant excess of events over the expected background. At $90\%$ confidence level, this corresponds to $2.8$ events~\cite{Bramante:2016rdh}. }

%and $1.4-10.6~{\rm KeV}$ for the electron recoil.$\text{Xe}^{132}_{54}$

%%%%%%%%%%%%%%%%%%%%%%%%%%%%%%%%%%%%%%%%%%%%%%%%%%%%%%%%%%%%%%%%%%%%%%%%%%%%%%%%%%%%%%%%%%%%%%%%%%%%%%%%%%%%%%%%%%%%%%%%%%%%%%%%%%%%%%%%%%%%%%%%%%%%%%%%%%%%%%%%%%%%%%%%%%%%%%%%%%%%%%%%%%%%%%%%%%%%%%%%%%%%%%%%%%%%%%%%%%%%%%%%%%%%%%%%%%%%%%%%%%%%%%%%%%%%%%%%%%%%%%%%%%%%%%%%%%%%%%%%%%%%%%%%%%%%%%%%%%%%%%%%%%
\bibliographystyle{utcaps}
\bibliography{References}

%% Bibliography
%\bibliographystyle{utcaps} 	% arXiv hyperlinks, preserves caps in title
%\bibliographystyle{utphys} 	% arXiv hyperlinks
%\bibliography{DM_self_Immo_collapse}

\end{document}